%% file: 00main.tex
\documentclass[trackchanges,twocolumn]{aastex7}
\usepackage{graphicx}
\input LaTeX-macros.tex
\usepackage[utf8]{inputenc}
\usepackage{xspace}
\usepackage{natbib}
\usepackage{array}
\usepackage{tabularx}
\newcolumntype{Y}{>{\raggedright\arraybackslash}p{4cm}}
\newcolumntype{A}{>{\centering\arraybackslash}p{1cm}}
\newcolumntype{B}{>{\centering\arraybackslash}p{2cm}}
\newcolumntype{E}{>{\centering\arraybackslash}p{2.1cm}}
\newcolumntype{C}{>{\centering\arraybackslash}p{3cm}}
\newcolumntype{D}{>{\centering\arraybackslash}p{3.5cm}}
\newcommand{\rr}[1]{{#1}}

\usepackage{makecell}

\begin{document}

\title{Properties of \water masers and their associated sources in Sagittarius B2}

\author[0000-0002-0533-8575]{Nazar Budaiev}
\affiliation{Department of Astronomy, University of Florida, P.O. Box 112055, Gainesville, FL 32611}
\email[show]{nbudaiev@ufl.edu}  
\author[0000-0001-6431-9633]{Adam Ginsburg}
\affiliation{Department of Astronomy, University of Florida, P.O. Box 112055, Gainesville, FL 32611}
\email{adamginsburg@ufl.edu}

\author[0000-0002-2542-7743]{Ciriaco Goddi} 
\affil{
Universidade de S{\~a}o Paulo, Instituto de Astronomia, Geof{\`i}sica e Ci{\^e}ncias Atmosf{\'e}ricas, Departamento de Astronomia, S{\~a}o Paulo, SP 05508-090, Brazil}
\affiliation{Dipartimento di Fisica, Universit\'a degli Studi di Cagliari, SP Monserrato-Sestu km 0.7, I-09042 Monserrato,  Italy}
\affiliation{INAF - Osservatorio Astronomico di Cagliari, via della Scienza 5, I-09047 Selargius (CA), Italy}
\email{cgoddi@gmail.com}

\author[0000-0002-3078-9482]{\'Alvaro S\'anchez-Monge}
\affiliation{Institut de Ci\`encies de l'Espai (ICE, CSIC), Can Magrans s/n, E-08193, Bellaterra, Barcelona, Spain}
\affiliation{Institut d'Estudis Espacials de Catalunya (IEEC), Barcelona, Spain}
\email{asanchez@ice.csic.es}

\author[0000-0002-1730-8832]{Anika Schmiedeke}
\affiliation{Green Bank Observatory, PO Box 2, Green Bank, WV 24944, USA}
\email{ajschmiedeke@nrao.edu}

\author[0000-0003-0416-4830]{Desmond Jeff}
\affiliation{Department of Astronomy, University of Florida, P.O. Box 112055, Gainesville, FL 32611}
\affiliation{National Radio Astronomy Observatory (NRAO), 520 Edgemont Road, Charlottesville, VA 22903, USA}
\email{astrosynth1867@protonmail.com}

\author[0000-0003-2141-5689]{Peter Schilke}
\affiliation{I. Physikalisches Institut der Universit\"at zu K\"oln,
Z{\"u}lpicher Str. 77, 50937 K\"oln, Germany}
\email{schilke@ph1.uni-koeln.de}

\author[0000-0003-3115-9359]{Christopher De Pree}
\affiliation{National Radio Astronomy Observatory (NRAO), 520 Edgemont Road, Charlottesville, VA 22903, USA}
\email{cdepree@nrao.edu}

\begin{abstract}

We present high-resolution Karl G. Jansky Very Large Array observations of the 22 GHz \water maser line in the extended Sagittarius B2 cloud. We detect 499 \water masers across the observed velocities between -39 and 172 \kms. 
To investigate the nature of the masers, we analyze their spatial distribution and cross-match with catalogs of \hii\ regions and protostellar cores. 
62\% of masers are associated with protostellar cores and 32\% with \hii regions. The nature of the remaining 6\% of sources was not established, but is likely associated with protostellar cores. 
Based on the spatial extent of the groups of masers, we classify them as either outflow-associated or young stellar object (YSO)-associated. We identify 144 unique sites of maser emission: 23 are associated with \hii regions and 94 with protostellar cores, of which 33 are associated with protostellar outflows and 18 with YSOs.

The outflow-associated \water maser emission is confined to within $<2000$ au of the central continuum source, despite  shocked \sio emission extending over tens of thousands of au. 
The YSO-associated masers show a lack of detections at $5 < V_{rel} < 30\kms$, which we suggest may be due to maser self-absorption. 
We show how \water masers trace the large-scale material flow in Sgr B2 N (North) also seen in \sio and mm continuum emission. 
Finally, we find that protostellar cores with associated \water masers tend to have brighter 3 mm continuum emission on average, although there is no strong correlation between maser brightness and continuum flux.  
\end{abstract}

\section{Introduction} \label{sec:intro}

The Central Molecular Zone (CMZ) hosts some of the most extreme star-forming conditions in the Galaxy, with elevated gas temperatures, high turbulence, and strong magnetic fields \citep{Henshaw2022}.
Within the CMZ, Sagittarius B2 (Sgr B2) is the most massive and active star-forming complex. Located approximately 100 pc from the Galactic center \citep{Reid2009}, Sgr B2 contains a network of star-forming protoclusters: North (N), Main (M), South (S), and Deep South (DS), with N and M being the most active centers of massive star formation.
Sgr B2 is known for its high star-formation rate \citep[$\sim$0.04 \msun$\yr^{-1}$, 10\% of the star formation in the CMZ;][]{Ginsburg2018}. It contains over 50 \hii regions \citep{Meng2022}, close to a thousand protostellar cores \citep[M. Daley in prep.]{Ginsburg2018,Budaiev2024}, and over a dozen hot cores \citep{Bonfand2019, Jeff2024}.
The combination of extreme environmental conditions and clustered star formation makes Sgr B2 a compelling analog for star formation at during the peak epoch of cosmic star formation \citep[$z \sim 2$;][]{Madau2014}.

Water maser emission at 22.2350798\ghz has long served as a tracer of massive star formation. These masers typically arise from shocks within protostellar outflows, disk winds, or ionization fronts \citep{deJong1973,elitzur1992,liljestrom2000,Goddi2005,Moscadelli2019}, and have been used extensively to trace kinematics and proper motions in embedded regions \citep[e.g.,][]{Goddi2006,Asanok2023, Zhang2024}.
However, the inability to resolve the immediate environments of individual massive protostars limited the physical interpretation of maser observations.
With the advent of Atacama Large Millimeter/submillimeter Array (ALMA) and James Webb Space Telescope (JWST), it is now possible to resolve forming stars and their surroundings within massive molecular clouds \citep{Budaiev2024, Zhang2025, Xu2025, Crowe2025}. The high-resolution multi-wavelength observations enable us to revisit existing maser data in the context of resolved feedback, infall, and dynamical interactions within the forming star clusters.

We report high-resolution Karl G. Jansky Very Large Array (VLA) observations of \water masers in the Sgr B2 cloud. In Section \ref{sec:data}, we present the VLA observations and describe the pipeline rerun and imaging procedures, including self-calibration. We then describe the source extraction methods in Section \ref{sec:methods}. In Section \ref{sec:analysis}, we compare our catalog with the previous observations of \water masers in the region and cross-match the detections with the existing catalogs of protostellar cores and \hii regions. 
Finally, in Section \ref{sec:discussion} we investigate the spatial distribution and clustering of masers to determine their nature and analyze the propeties of the associated sources.
The summary of the work is provided in Section \ref{sec:conclusions}. 

\section{Observations and Data Reduction} \label{sec:data}

\subsection{Observation description}
The observations of the 22.2350798\ghz \water maser line towards Sgr B2 are a part of VLA project 18A-229 (PI: A. Ginsburg). The data were collected in two observing sessions on March 28th and 29th, 2018 using A configuration. The three fields covered a 10.6 $\times $ 6\pc field of view extending from Sgr B2 N to DS (Fig. \ref{fig:overview_flux}). The observations were centered on 22.2330969\ghz to account for the velocity of the cloud ($V_{\mathrm{LSR}}=60\kms$), with a total bandwidth of 16 MHz and a channel width of 0.07\kms.  The beam size is $0.\arcsec16\times0.\arcsec06$ \citep[$1300 \times 500$ au at Sgr B2's distance of 8.277\kpc;][]{GRAVITY2022}. The total on-source time is just under 42 minutes for each of the three fields.
The observations used 1331+305=3C286 as the flux calibrator, J1733-1304 as the bandpass calibrator, and J1744-3116 \footnote{As reported in \cite{Budaiev2024}, VLA phase calibrator J1744-3116, which is used for many observations of Sgr B2, has wrong coordinates recorded in the VLA catalogs when compared to other catalogs, such as \citep{Lanyi2010}.} as the phase calibrator.
We note that the labels for fields ``MS" (Main-South) and ``SDS" (South-Deep South) are swapped in the observation logs, but only for K-band observations. The ``MN" (Main-North) field is labeled correctly. In this paper, we refer to fields based on what was observed, but the reduction code primarily uses the observational log's labels.

\subsection{Pipeline rerun}
Due to the extreme brightness of the observed masers (up to 200\jy) the standard pipeline would flag most of the signal as radio frequency interference. We reran the VLA pipeline (version 6.4.1.12) with manually specified continuum (signal-free) channels and Hanning smoothing turned off.

\subsection{Antenna flagging, channel flagging, scan flagging}
Upon inspection of amplitude vs. frequency plots, we identified problematic antennas ea18 and ea05, which we subsequently flagged in both measurement sets. Upon further evaluation of the data, we discovered that the fourth scan for each field in the measurement set observed on March 28th produces images with much higher noise and more spatially spread out flux, which could not be addressed with self-calibration. Thus, we excluded the fourth scan for both measurement sets.

\subsection{Spatial offset between two measurement sets}
Upon inspection of the preliminary images from the two measurement sets, we discovered a spatial offset in emission. The maser locations were spatially offset by $\sim$0.\arcsec1 in all fields between the two measurement sets, primarily in the North-South direction.
There are several possible explanations for this offset, but the true source could not be established.
The two measurement sets were observed only one day apart; thus, the offset cannot be attributed to proper motion.

In addition, the quality of the phase solutions was not uniform over different scans, with the earlier scans having worse solutions. This was particularly the case for the measurement set observed on Match 28th. As the result, the peak of the emission is not aligned between scans, which could have contributed to the offset.

Finally, some of the offset can be attributed to VLA's absolute pointing accuracy. The best-case pointing accuracy for the observing configuration is $\sim$10 mas. However, the observing conditions, especially the low elevation of Sgr B2, can significantly impact the final pointing accuracy.

We imaged both measurement sets with a manual input model such that the data converge to the same location and aligned the image with high-resolution ALMA continuum from \cite{Budaiev2024}. The details of this process are explained in Sections \ref{sec:self-calibration}, \ref{sec:cube_imaging}, and \ref{sec:wcs_shift}.

\begin{figure*}
    \centering
    \includegraphics[width=1\linewidth]{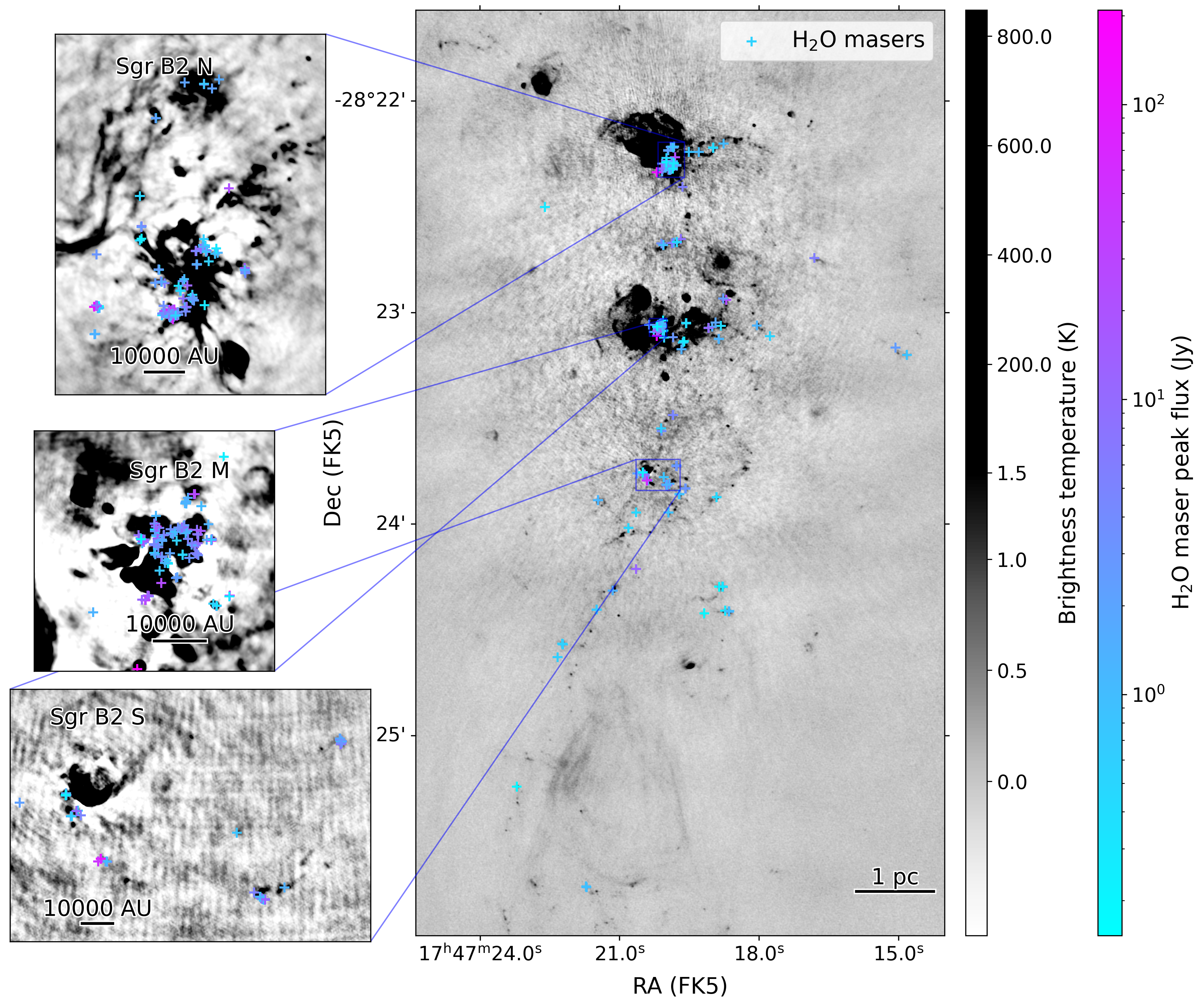}
    \caption{Locations and fluxes of \water masers in Sgr B2. The background is 3 mm ALMA continuum from \cite{Ginsburg2018} The zoom-ins show masers in the Sgr B2 N(orth), M(ain), and S(outh) star forming clusters plotted on top of the high-resolution 3 mm continuum from \cite{Budaiev2024}. The shown field-of-view corresponds to the imaged area with the VLA.}
    \label{fig:overview_flux}
\end{figure*}

\subsection{Maser self-calibration} \label{sec:self-calibration}

We self-calibrated the data on the channel with the brightest maser emission in each field. 
Considering the spatial offset between the measurement sets and the scans described in the previous section, we use a manually generated input model for the first round of self-calibration.
The model consists of a single pixel located in the average position of the brightest maser between two measurement sets.
As a result, all scans and measurement sets converge to a single location, so the measurement sets can be imaged together.

\begin{figure}
    \centering
    \includegraphics[width=0.49\linewidth]{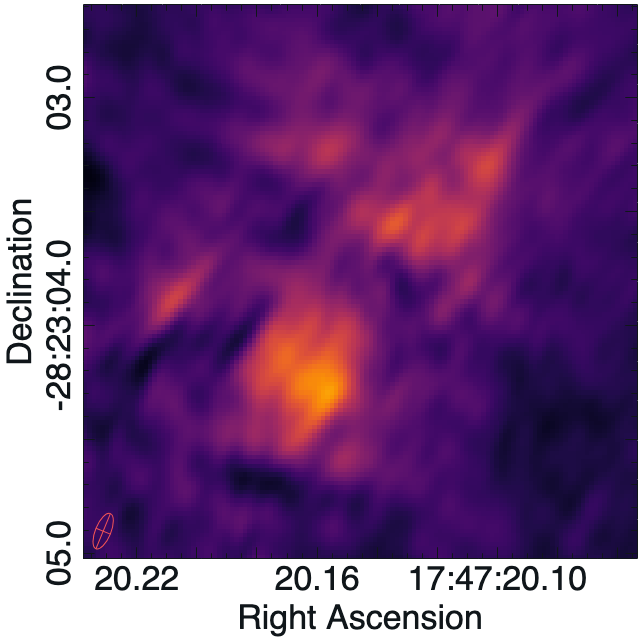}
    \includegraphics[width=0.49\linewidth]{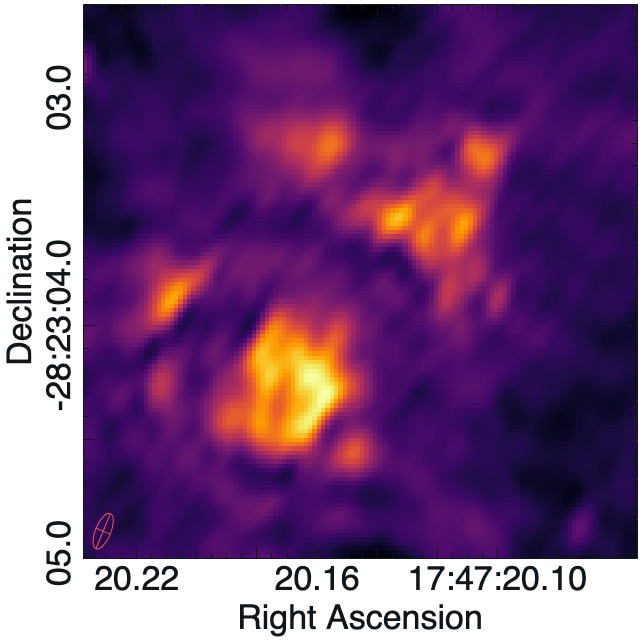}
    \caption{A zoom-in on Sgr B2 M highlighting the differences of pre-self-calibration (left) and post-self-calibration (right) continuum image. Both images use the same color scale. The self-calibration solutions from \water maser spectral windows were applied to all continuum spectral windows. }
    \label{fig:MN_calibration}
\end{figure}

We applied the self-calibration solution tables to the continuum spectral windows to produce a 22 GHz continuum map of Sgr B2 (see Appendix \ref{app:continuum_image} and Figure \ref{fig:continuum_22GHz}). The continuum image is later used to astrometrically align these observations to ALMA observations of the same region (see Section \ref{sec:wcs_shift}).
Figure \ref{fig:MN_calibration} demonstrates the improvement gained with self-calibration: better phase solutions result in more concentrated emission. 
The fully reduced continuum files are available on Zenodo via doi:\href{https://doi.org/10.5281/zenodo.15747715}{10.5281/zenodo.15747715}.

\subsection{Cube imaging} \label{sec:cube_imaging}

Computational constraints were the driving factor in the selection of most of the imaging parameters.
We downselect the cube for imaging: we exclude the first and last 32 channels. We inspect these channels separately to ensure that there are no masers present. We also restrict the imaged area to 6 $\times$ 6\pc for each field. At the edges of the square field of view, the primary-beam response ranges between 0.33 for the inscribed circle radius and 0.08 for the circumscribed circle radius. With these restrictions the size of one cube for each field was $\sim$700 GB.

We use CASA's \citep[version 6.6.0.2,][]{McMullin2007} \textit{automultithresh} automatic masking algorithm to generate cleaning masks. Due to the volume of data, it is impractical to perform interactive cleaning. However, \textit{automultithresh} was developed on ALMA data and does not perform as well with the data in hand. For example, after the first $\sim$10 individual masks were created in a channel, the algorithm would begin masking artifacts and the number of masks would rise rapidly. Masking and consequently cleaning artifacts results in inaccurate flux measurements. Thus, we adjusted \textit{automultithresh} parameters until there were no more than 10 masks per channel. While some of the fainter sources in a channel may not get masked, the brightest ones, responsible for the most artifacts, do get masked.   
These masks are used for the remainder of the cleaning process. We then clean each cube to several threshold levels (2\jy, 0.5\jy, 0.1\jy) and inspect the results for divergence or overcleaning when each cleaning threshold is reached.

\begin{figure*}
    \centering
    \includegraphics[width=\textwidth]{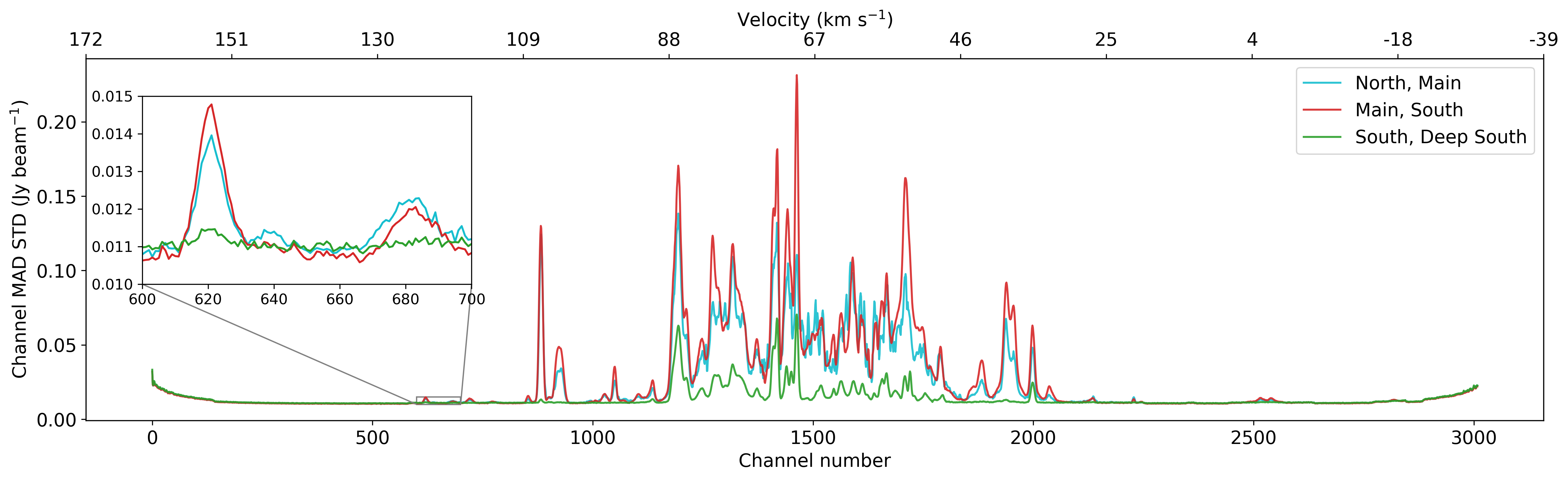}
    \caption{Median-absolute-deviation (MAD)-based standard deviation for each channel of the cube. The sensitivity is artifact-noise limited for a third of the channels, where most of the masers are found. }
    \label{fig:noise}
\end{figure*}

The final imaged velocity range is from -39.14\kms and 172.06\kms, with a spectral resolution of 0.07\kms.
The per-channel median-absolute-deviation-based (MAD) standard deviation for each channel, which is used as a noise estimate, is shown in Figure \ref{fig:noise}.
The imaging process took approximately one month of computational time for each of the three cubes using the University of Florida Research Computing machine HiPerGator.

\subsection{WCS shift} \label{sec:wcs_shift}
After successfully self-calibrating on a bright maser to an arbitrary location, we perform spatial alignment. We applied the resulting self-calibration tables on the continuum spectral windows of the observations to obtain a well-behaved 1.3 cm continuum image. Then, we use the phase cross-correlation algorithm from the \texttt{scikit-image} python package \citep{scikit-image}. We then cross-correlate the 3 mm ALMA continuum image \citep{Budaiev2024} with the 1.3 cm continuum image. The two data sets have similar resolutions. The ALMA observations happened 8 months prior to the VLA observations, thus any expected deviations due to proper motions are negligible. ALMA's absolute astrometry for long-baseline observations is $\sim$10\% of the beam size, 70 au in our case, and is better than that of the VLA.

The northernmost field observed by VLA has a similar field-of-view to that of the ALMA 3 mm observations. We select five regions of bright, extended emission to perform cross-correlation (Sgr B2 M, two extended \hii regions near Sgr B2 N, and two extended \hii regions near Sgr B2 M). We then take the average of the results and propagate the errors; the final cross-correlation error is 0.19 pixels (3.6 mas). 

We bootstrap our WCS solutions from the northernmost field to the remaining two fields by spatially fitting the locations of masers. We identify the four brightest masers within the spatial overlap between all fields.
For each maser, we select the channel with the brightest emission and fit a 2D Gaussian to each maser. We then calculate the shift in pixels. Between the four masers, the calculated coordinate shift variation was no larger than 3\% of the offset distance. We take the average offset from the four measurements and apply it to the remaining two fields by editing the `CRVAL' in the FITS headers of the maser cubes. The propagated error is 0.2 pixels (4 mas). 


The best-case absolute positional accuracy of our observations is $\sim$0.3 pixels (6 mas, 50 au). However, we caution the reader that the reported uncertainty is not necessarily representative of true uncertainty (e.g. cross-correlation does not account for the changes in physical properties of the observed objects in the 22 and 100 GHz images). 

\section{Methods}\label{sec:methods}

\subsection{Source extraction}\label{sec:source_extraction}

We utilized \texttt{astrodendro} \citep{Rosolowsky2008}, a dendrogram-based python package to extract sources.
Due to the large number of extremely bright ($>$ 100 \jy) masers, the per-channel noise varies significantly throughout the cube. 
Since \texttt{astrodendro} is not designed to work with varying noise, we implemented an approach to split the cube into smaller cube chunks with similar noise. 

First, we calculated the standard deviation based on the MAD for each channel, which should be representative of the noise even in channels with several bright masers. Then, we defined noise bins that will determine how the cube will be split: the first noise bin was defined as the lowest noise in the cube and twice that value. The next noise bin starts at the previous value and ends at twice that, and so on. 
This ensures that the assumed noise for the source extraction algorithm will be different by no more than a factor of two from the actual noise.
Then, all consecutive channels that fall into the same noise bin are grouped into chunks. After the channel chunks are determined, we expand each chunk range by one on each side such that all chunks now overlap by two channels. This is done to simplify the source extraction pipeline later on, as the dendrogram will pick up any source that has rising spectrum within a chunk but peaks outside outside of the relevant channel chunk.

Finally, we run \texttt{astrodendro} on each of the channel chunks on the non-primary-beam-corrected images. After preliminary testing, we determined that the three imaged fields require slightly different parameters to achieve the best balance between the number of true positives and false positives. We fix the \textit{num\_pix} at 45 (the number of pixels per beam), varied the \textit{min\_value} parameter, and set the \textit{min\_delta} to 0.7 the \textit{min\_value} parameter). We start at \textit{min\_value} equal to four times the highest noise in each chunk (equivalent to eight times the lowest noise) and gradually increased the threshold until the number of false positives was less than number of true positives with a by-eye inspection. The final \textit{min\_value} parameters for each field were: 16 for MN, 14 for MS, and 10 for SDS.

Next, we combine the dendrogram-generated catalogs from each of the chunks. 
Since the neighboring channel chunks overlap by two channels, we remove all detections in the edge channels for each of the chunks.
This ensures that the rising emission profile of the line that peaks in the following chunk is not cataloged as a source in the preceding chunk.
This removed all false positives detections caused by any source that lies outside, but close to, the channel chunk.
We construct the final catalog by combining the merged catalogs from each of the fields. If a source is present in multiple fields, we prioritize the distance to the center of the imaged field as the selection criteria.

\subsection{Completeness estimation}

To estimate the noise, we used the MAD standard deviation for each channel from Figure \ref{fig:noise}. The typical noise in a signal-free channel is 11\mjyb. The presence of bright emission results in a higher per-channel noise, primarily between 50 and 90 \kms.  The brightest masers are abundant in Sgr B2 N and M, thus the field covering Sgr B2 S and DS stands out with the lower noise.

The sensitivity of our data varies radially (due to radio beam sensitivity drop-off), varies with frequency (due to some channels containing bright emission that cannot be fully cleaned), and varies spatially (due to local artifacts within a single channel). The varying sensitivity complicates the completeness estimation. 
We estimated the completeness based on the noise statistics of the extracted sources. This is important because if the majority of the sources were extracted near channel 1450, where the noise is $\sim$100\mjyb, then the completeness value is much higher than if the majority of detections came from channels with $\sigma = 11\mjyb$. 

\begin{figure}
    \centering
    \includegraphics[width=1\linewidth]{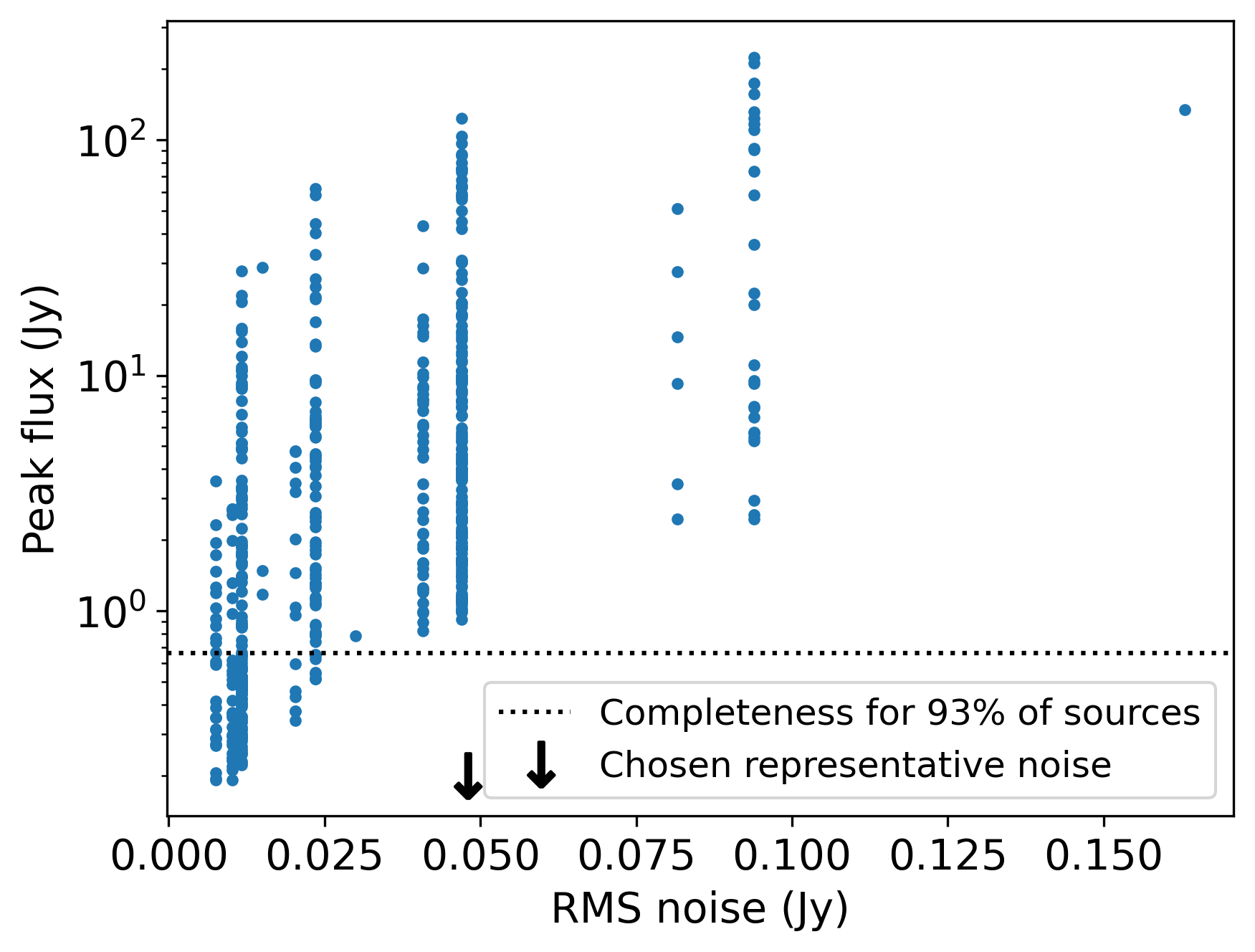}
    \caption{A distribution of fluxes of the extracted sources and the noises of the channels they were extracted from. 93\% of sources (to the right of the arrow) were extracted from channels with noises below 0.046\jyb. This value is chosen as the noise for the completeness calculation. Multiplying this noise by 14, the average signal-to-noise ratio used during source extraction, results in the final catalog completeness of 0.66\jyb, shown as a horizontal dashed line.}
    \label{fig:completeness}
\end{figure}

We show the distribution of channel noises from which the sources were extracted and the sources' brightnesses in Figure \ref{fig:completeness}.  93\% of sources were extracted from channels with noise below 0.046\jyb. Thus, we choose 0.046 \jyb as our representative noise for the completeness calculation. 
As described in Section \ref{sec:source_extraction}, the signal-to-noise threshold was different for each field: 16 for MN, 14 for MS, and 10 for SDS. We choose $14\sigma$ as a representative completeness for our sample. Thus, we estimate our catalog completeness to be 0.66\jyb. While several assumptions were made to obtain this catalog completeness, it should be representative for the majority of the sample. 

\begin{figure}
    \centering
    \includegraphics[width=0.99\linewidth]{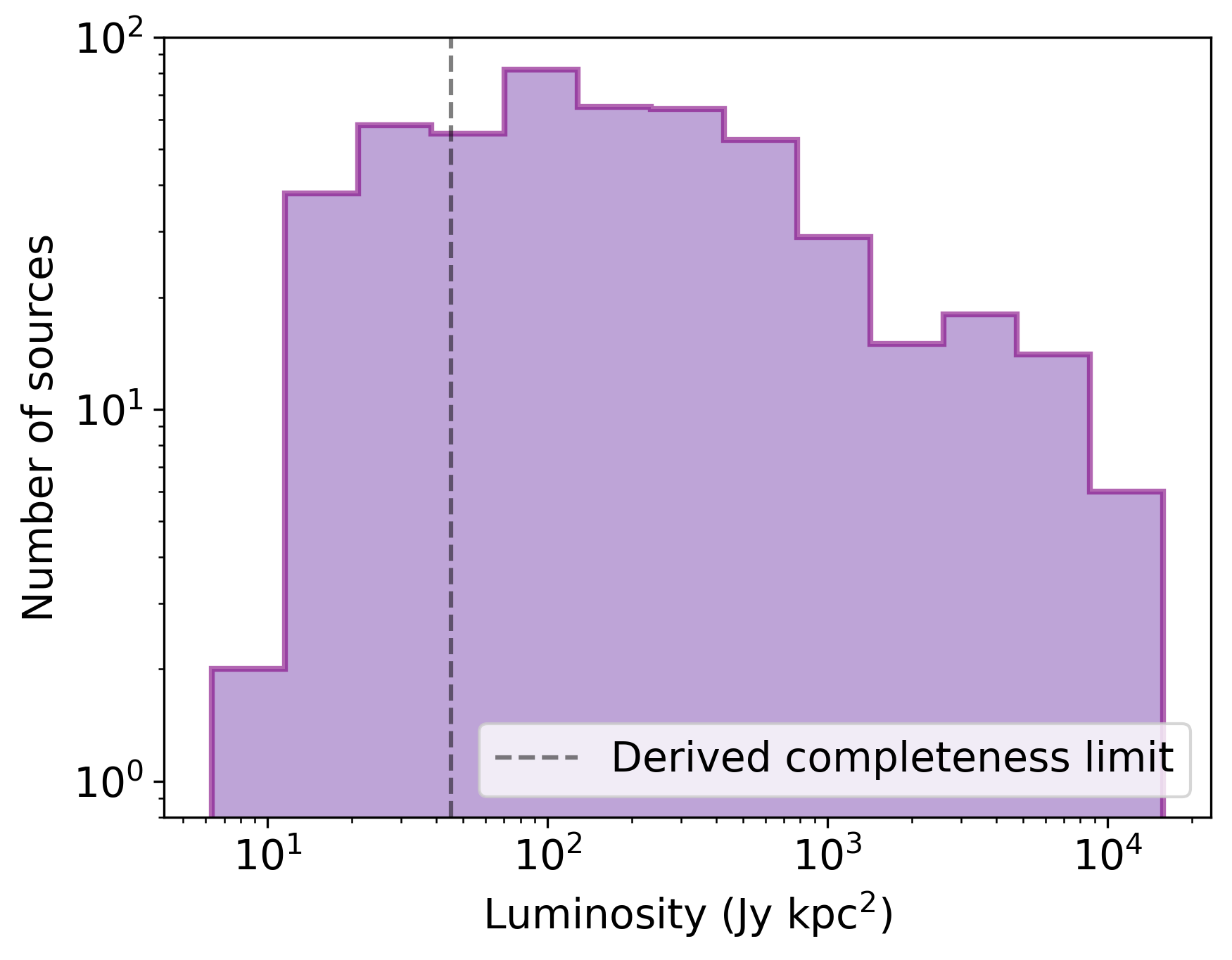}
    \caption{Isotropic luminosities of \water masers in Sgr B2. Each spatial and spectral detection is shown separately}
    \label{fig:luminosities}
\end{figure}

We report the isotropic \water maser luminosities for each detection assuming d=8.277 \kpc \citep{GRAVITY2022}. The histogram of the luminosities is shown in Figure \ref{fig:luminosities}.

\subsection{Spatial fitting, spectral fitting with pyspeckit}
We fit the dendrogram-detected sources spatially and spectrally to increase the precision of our measurements. We use the location and flux of the brightest pixel in each source as input parameters for fitting. 
First, we fit sources spatially with a 2D Gaussian fitting in the channel with brightest emission. We visually inspect each detection and reject any spurious detections caused by beam sidelobes or imaging artifacts. Only 8 sources out of the 499 had unsatisfactory fits and are marked as such in the catalog (see Table \ref{tab:catalog}). 
The median fit uncertainty is 0.1 pixels (2 mas) in Right Ascension and 0.2 pixels (4 mas) in Declination. The relative positional uncertainty in the catalog is $<$5 mas ($<$50 au at Sgr B2's distance).

In addition, we inspect the spectra of each source. In rare occasions, \textit{automultithresh} masking would result in unphysical changes in brightness in the channel preceding the mask. Consequently, \texttt{astrodendro} would identify these as sources. We reject these detections at this step.

With the selected \textit{min\_delta} parameter, \texttt{astrodendro} was successful at spatial de-blending for the majority of clustered sources. While some sources were merged and thus identified as a single detection, the conclusions drawn in this work are unaffected.

We then fit the sources spectrally using \texttt{pyspeckit} package \citep{pyspeckit}. We center on each detection with a range of 140 channels. To make the automated fitting process more robust, we add additional guesses to the spectra: we search the $30\times30\times140$ subcube around the cataloged detection for other detections, and us their extracted parameters as initial guesses. We us 0.1\kms as the FWHM guess, similar to the channel width. Again, we inspect each fit and refit the unsuccessful fits manually using the GUI interface available in \texttt{pyspeckit}. The full width at half maximum for each source is $<$0.1\kms (1.5 channels), indicating that all sources are unresolved spectrally.

\begin{figure*}
    \centering
    \includegraphics[width=1\linewidth]{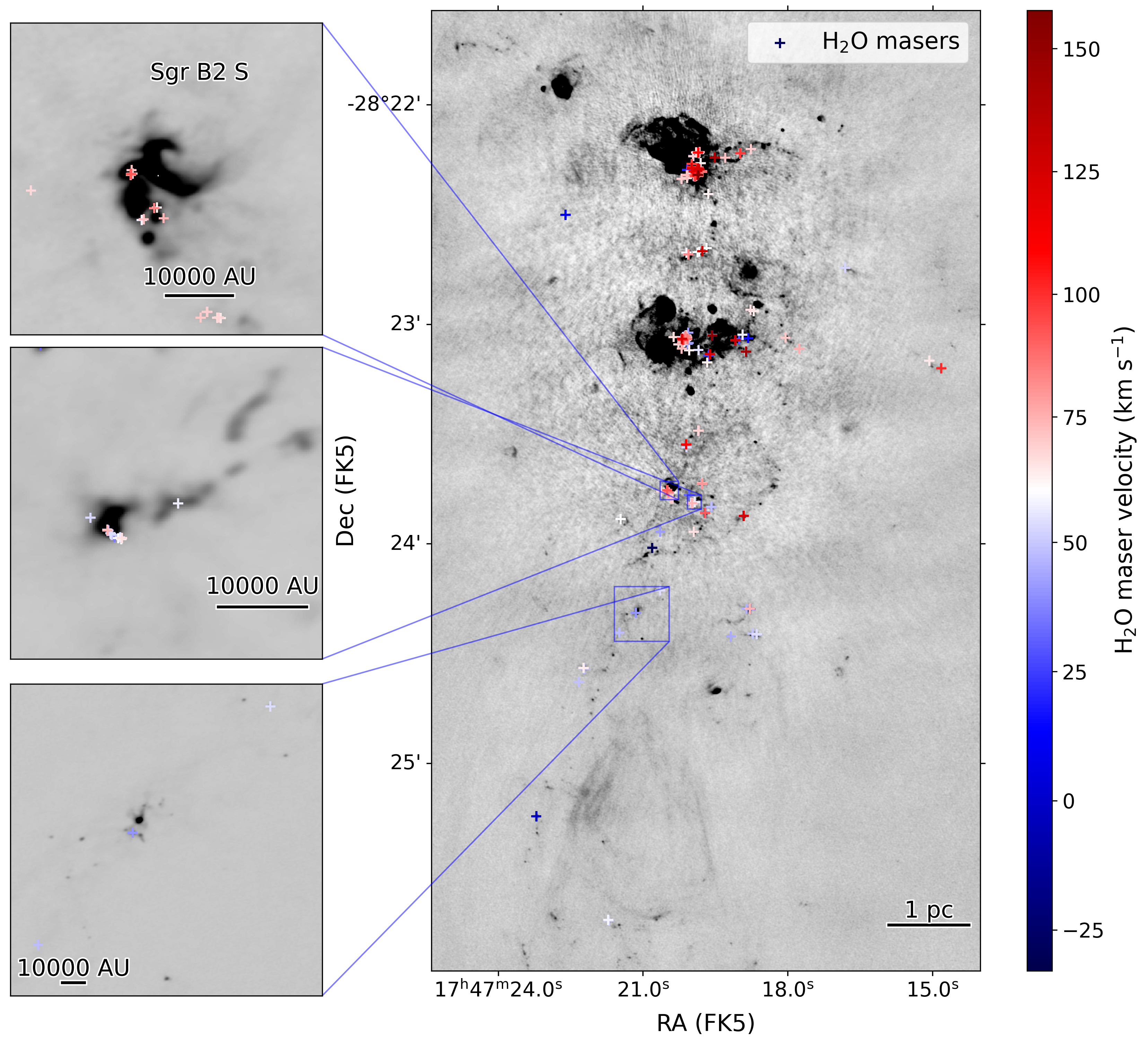}
    \caption{Locations and velocities of \water masers in Sgr B2. The background is 3 mm ALMA continuum from \cite{Ginsburg2018} The zoom-ins show masers in Sgr B2 S(outh) and D(eep)S(south) with high-resolution 1 mm continuum (ALMA PID:2017.1.00114.S).}
    \label{fig:overview_velocities}
\end{figure*}

The final catalog contains 499 \water maser detections in the Sgr B2 molecular cloud shown in Figures \ref{fig:overview_flux} and \ref{fig:overview_velocities}. The sample from the catalog is shown in Table \ref{tab:catalog}. The full catalog is available in machine-readable format.

\section{Analysis}\label{sec:analysis}

\begin{figure}
    \centering
    \includegraphics[width=0.99\linewidth]{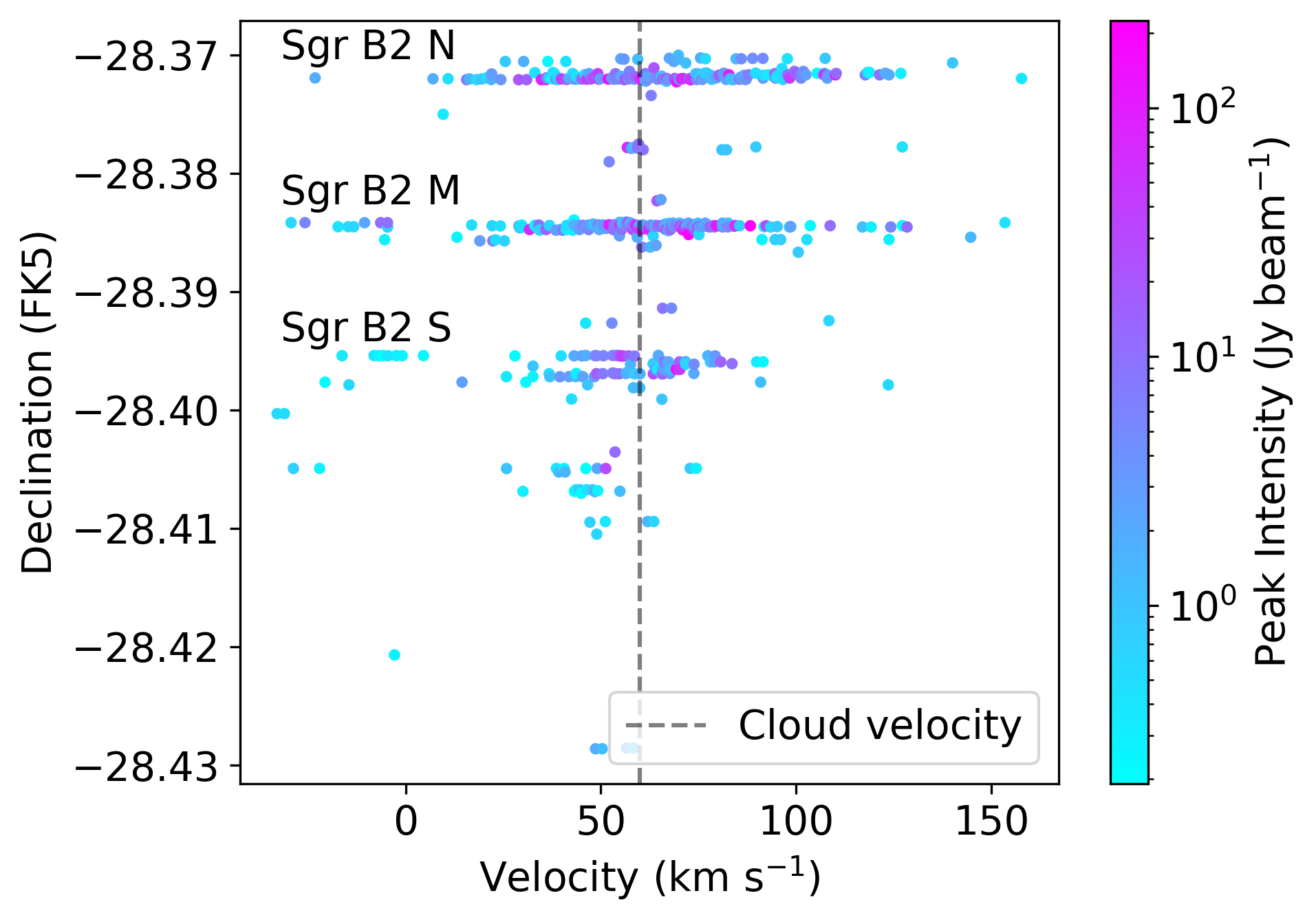}
    \caption{\water maser velocities plotted against Declination in Sgr B2. The horizontal features, from north to south, correspond to Sgr B2 N, Sgr B2 M, and Sgr B2 S with the surrounding hot cores. The maser velocities have a similar distribution in N and M, ranging between -25 and 125\kms. 
    The vertical dashed line indicates the cloud's relative velocity.}
    \label{fig:velocity_vs_dec}
\end{figure}

\subsection{Comparison with McGrath 2004 catalog}
The previous observations of \water masers in the extended regions of Sgr B2 were described in \cite{McGrath2004}. Their velocity coverage is between -40 and 120\kms, with a spectral resolution of 0.66\kms, compared to ours of between -39 and 172\kms at 0.07\kms. \cite{McGrath2004} report the typical rms noise in a line-free channel as 1\jyb, compared to 0.011 \jyb in our data. Accounting for their slightly larger beam, our absolute sensitivity is 12 times better than their rms noise. Spatially, our observations extend further in the negative declination, covering the younger part of Sgr B2 (DS). 
With better sensitivity, higher spatial resolution, and larger spectral coverage, we detect $\sim$2.5 times more \water masers in the overlapping part of the fields of view.

Their observations were conducted in June 1998, resulting in a $\sim$20 year baseline between the observations. 
\water masers are known to vary on much shorter timescales \citep[e.g., months; see][]{Zhang2024}. As expected, we find significant changes in source brightnesses. Due to the high source density and the large time baseline between two observations, it is impractical to perform variability analysis.

\rr{The distribution of fluxes above the common completeness limit is similar between the two catalogs. This similarity could be coincidental with two observations having similar flux distributions by chance. Alternatively, and more likely, it may suggest that the overall flux distribution remains relatively stable over time, with only a few bright outliers, caught near the peak of their variability cycles, dominating the detectable maser emission. Thus, the population-level statistics shown in this work are likely to remain consistent over time. Follow up VLA observations of the region will show whether the underlying flux distribution persists across multiple epochs. 
}

\begin{figure}
    \centering
    \includegraphics[width=1\linewidth]{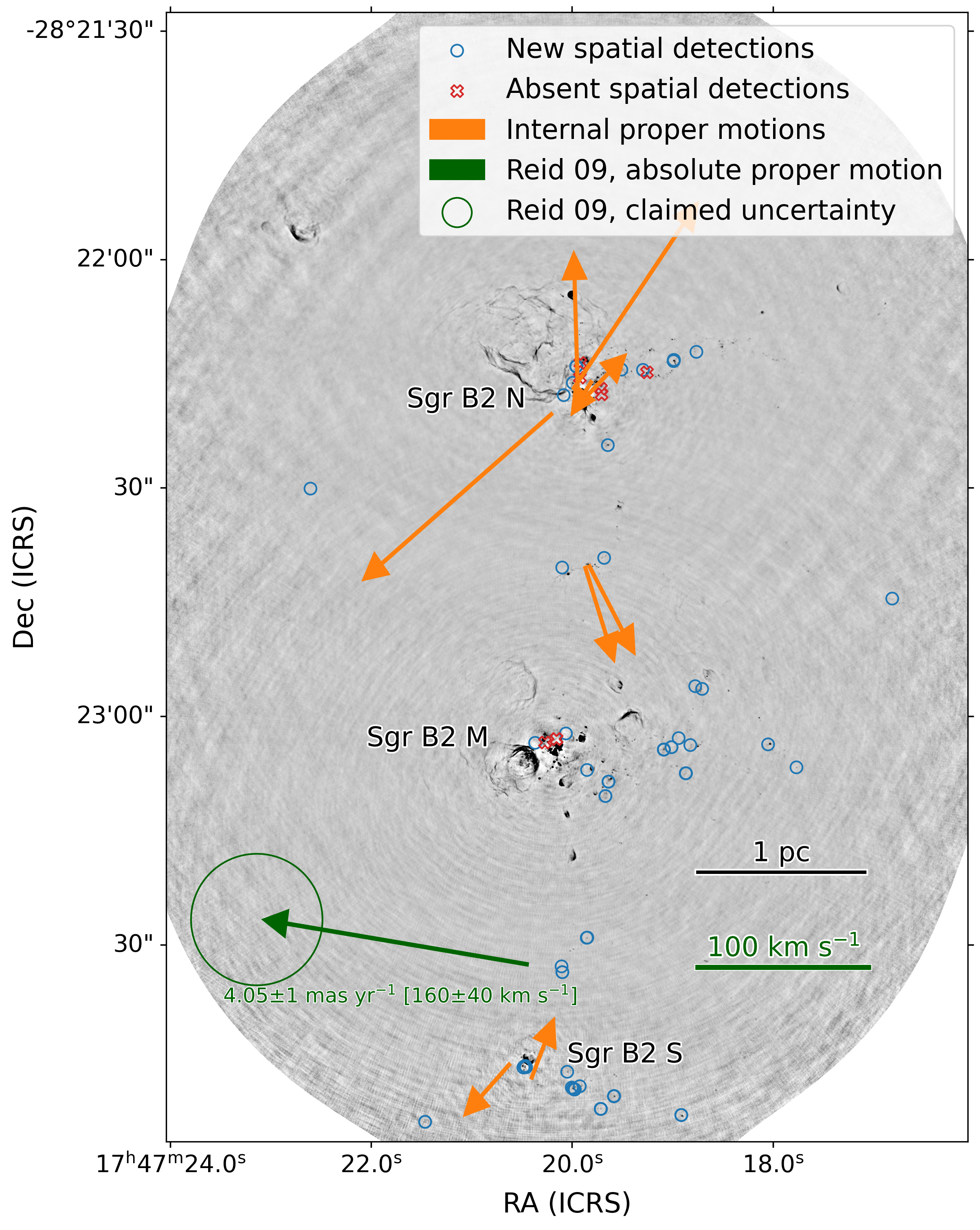}
    \caption{Locations of \water maser detections that do not have a spatial match between \cite{McGrath2004} and our catalogs within 0.5\arcsec ($\sim$4000 au). The pictured field-of-view is representative of that of \cite{McGrath2004}, thus all detections south of the figure (below Sgr B2 S) are new (see Figures \ref{fig:overview_flux} and \ref{fig:overview_velocities}). We find new detections both isolated from previously detected masers and clustered near previous detections. The missing detections are likely due to variability of the maser brightnesses.
    The \rr{green} arrow represents the absolute proper motion vector observed by \cite{Reid2009} (which includes the apparent motion of Sgr A*) and the green circle shows the reported uncertainty. The orange arrows show the internal proper motions of masers from this work. The relative length of the orange vectors compared to the \cite{Reid2009} uncertainty highlights that \water masers can be unreliable sources of proper motion measurements of Sgr B2, especially in small numbers. }
    \label{fig:overview_new_detections}
\end{figure}

We cross-match our catalog with the catalog from \cite{McGrath2004} and show the locations of \water masers without spatial matches within 0.\arcsec5 ($\sim$4000 au). The details of the cross-matching process are described in Section \ref{sec:proper-motion}. The cross-match search radius is several times larger than the resolution of either observation to highlight locations not previously associated with \water masers. Figure \ref{fig:overview_new_detections} shows the spatial distribution of the new and absent \water maser detections.

\subsection{Maser relative proper motion}\label{sec:proper-motion}

Upon comparing the spatial distributions of our catalog and that of \cite{McGrath2004}, we found a consistent shift present in all sources in the North-South direction. The systemic shift indicates a poor astrometric accuracy of one (or both) of the observations. 

We quantify the positional offset by cross-matching detections between the two catalogs. To ensure robust matches, we select sources that are isolated and have $<5\kms$ velocity difference between the two catalogs. Since a large fraction of \water masers are detected in groups, only nine sources meet the above criteria. 

For these matched sources, the average positional separation between our catalog and that of \cite{McGrath2004} is 0.\arcsec03 (250 au) in RA and -0.\arcsec12 (1000 au) in Dec. Over the 20-year baseline between observations, this offset implies a proper motion of the cloud of $\sim270\kms$, significantly larger than values suggested by previous observations and models \citep[e.g.][]{Kruijssen2015, Sormani2020, Takeshi2022}. 
\cite{McGrath2004} report an absolute astrometric accuracy of $\sim$0.\arcsec01 based on alignment of the centroids of continuum emission from \hii regions with those of \cite{Gaume1995}.
However, we were unable to access either continuum dataset to compare with ours.

Our VLA observations are astrometrically aligned to those of ALMA, which typically offer superior absolute positional accuracy. Given the lack of independent verification for the \cite{McGrath2004} absolute positions and the alignment of our data with ALMA, we conclude that the positions reported by \cite{McGrath2004} are likely offset from the true source locations.
While readers should be aware of the potential absolute astrometric uncertainties in both datasets, the conclusions in this work are based on comparisons with the ALMA continuum to which the current VLA observations were aligned, and thus are not affected by the absolute astrometric calibration.

\begin{figure}
    \centering
    \includegraphics[width=0.99\linewidth]{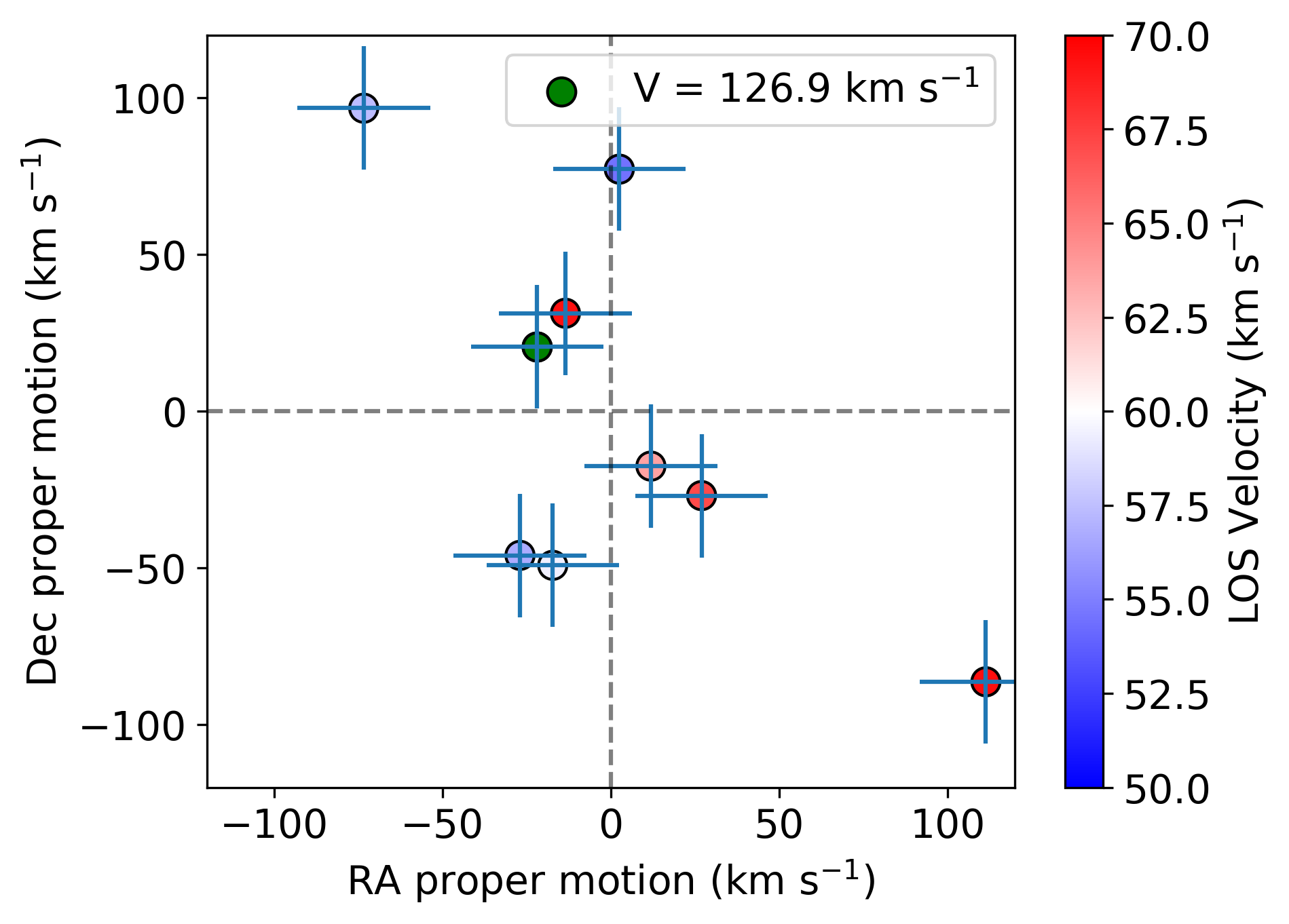}
    \caption{The measured source-to-source proper motions of \water masers in Sgr B2. The proper motions are measured over 20 year baseline between the observations in this paper and \cite{McGrath2004}. The theoretical velocity resolution is 12 \kms. However, due to a poor absolute astrometric match between the two \water maser catalogs and uncertainty in the correction procedures, this figure should be interpreted as proof of concept for measuring proper motions with VLA at \kpc distances rather than the true proper motion measurements. The proper motion vectors of the nine sources are shown as orange arrows in Figure \ref{fig:overview_new_detections}. The green circle is an outlier with a high LOS velocity.}
    \label{fig:proper_motion}
\end{figure}

We shift the \cite{McGrath2004} catalog by the calculated systemic offset. The relative, source-to-source proper motions for the nine cross-matched sources are shown in Figure \ref{fig:proper_motion}. Both \cite{McGrath2004} and our catalogs have relative source positional accuracy of 0.\arcsec005. However, \cite{McGrath2004} did not report their source extraction methods, so the validity of their uncertainty could not be verified. Thus, we choose 10 mas as a representative uncertainty for the error bars.

Two masers have proper motions over 100\kms, much higher than the typical 1D velocity dispersion of $\sim$12\kms. Such high proper motions are more consistent with the velocities we expect from outflows, oriented along the plane-of-sky. Indeed, for all but one source, the line-of-sight velocities are within 10\kms of the $V_{\mathrm{LSR}}$ of the cloud, which is consistent with the plane-of-sky outflows.

We note that such relative proper motions are proof-of-concept rather than robust measurements. \water masers are known to vary on much shorter timescales than the two-epoch baseline of 20 years. These results should be interpreted as part of a larger discussion on measuring cloud proper motion (see Section \ref{sec:PM_Reid}).

\subsection{Maser associations}\label{sec:associations}
We cross-match the \water maser detections with foreground stars, protostellar cores, and \hii regions. 
We use VLA- \citep{DePree1998, DePree2015, Meng2022} and ALMA-based \citep{Ginsburg2018, Budaiev2024} catalogs of \hii regions augmented with JWST MIRI data (N. Budaiev 2025, in prep). 
We associate a \water maser with an \hii region if it is either within the physical boundaries of an \hii region, or, for unresolved sources, is within one beam of the detection. 162 masers (32\%) are associated with \hii regions, with none associated with \hii regions larger than 0.025 pc (5000 au).

\sio masers are known to trace currently-forming, very massive stars and evolved massive stars. We identify evolved stars by cross-matching isolated \water masers with \sio masers from the large ALMA survey ACES (S. Longmore in prep).
As we have already identified \hii region-associated \sio masers, we exclude any \sio masers within 0.3 pc of known \hii regions.
Using a search radius of 0.1 pc, a single match is identified. 
We confirm the classification with JWST data (N. Budaiev 2025, in prep.) which also have a detection in all observed filters between 1.5 and 25 $\mu m$.
The source also has faint 3 mm emission, also typical for an evolved massive star.
Thus, we classify this maser as associated with an evolved massive star.

We then cross-match the \water masers with mm continuum data to find protostellar core associations. Having already classified massive evolved stars and \hii regions, we have minimized the risk of contamination. We utilize the $\sim$1000 au resolution catalog from \cite{Budaiev2024} and a $\sim$5000 au resolution catalog that covers the extended cloud \citep{Ginsburg2018}. 
If no match is found (or no data are available) within a 0.\arcsec3 ($\sim$2500 au) search radius of the higher-resolution data, we search within 0.\arcsec5 ($\sim$4500 au). In total, we find 259 sources associated with protostellar cores via cross-matching.
55 remaining \water masers have no matches between $\mu$m and cm wavelengths and are marked as ``no match" (see Table \ref{tab:catalog}).

\section{Discussion}\label{sec:discussion}

\subsection{Nature of water masers}\label{sec:nature}

To our knowledge, this is the first work that investigates the properties of \water masers and their associated star-forming objects within one site of massive star formation with statistically significant samples for different populations. Thus, the differences in the properties of different populations of \water masers can be attributed directly to their nature rather than the varying environment in the different star-forming regions.

\begin{figure}
    \centering
    \includegraphics[width=1\linewidth]{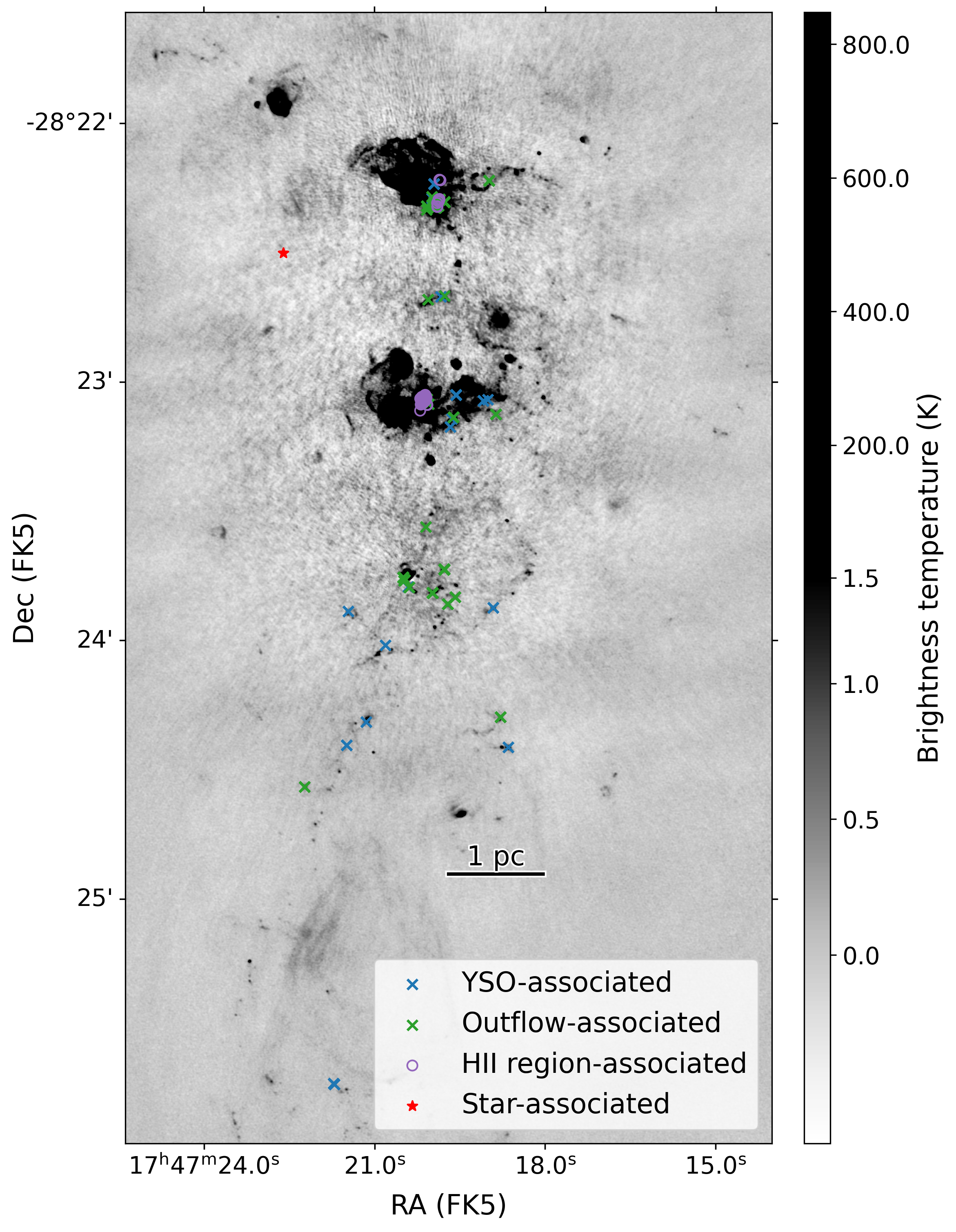}
    \caption{Overview of the spatial distribution of different populations of \water masers. No masers are associated with \hii regions larger than 5000 au. }
    \label{fig:overview_spatial}
\end{figure}


\subsubsection{Classifying masers around protostellar cores}
In Section \ref{sec:associations}, we associated \water masers with foreground stars, \hii regions, and protostellar cores. 
We now further classify the masers associated with protostellar cores and those that did not have any matches in the existing continuum catalogs into two groups: \water masers collisionally pumped by the YSO's outflow and \water masers attributed to different physical processes other than outflows (e.g. due to disk-winds, polar winds, radiative effects). 
In this work, we refer to them as ``outflow-associated" and ``YSO-associated," respectively. 
The main differentiating criteria between the two population is the spatial extent of the maser groups. The outflows can extend to thousands of aus and, correspondingly, so can the collisionally-pumped \water masers. Meanwhile, other physical processes are confined to few hundreds of au. Observations with VLBI or the ngVLA will allow us to distinguish between the different origins of YSO-associated \water masers through high spatial resolution or kinematic information. Since the ``no match" masers likely do have dust continuum associations that are below the sensitivity limit, we combine them with the protostellar core-associated masers for a total sample of 336 sources.

\begin{figure*}[htbp] 
    \centering
    \begin{minipage}{0.45\textwidth} 
        \centering
        \includegraphics[width=\linewidth]{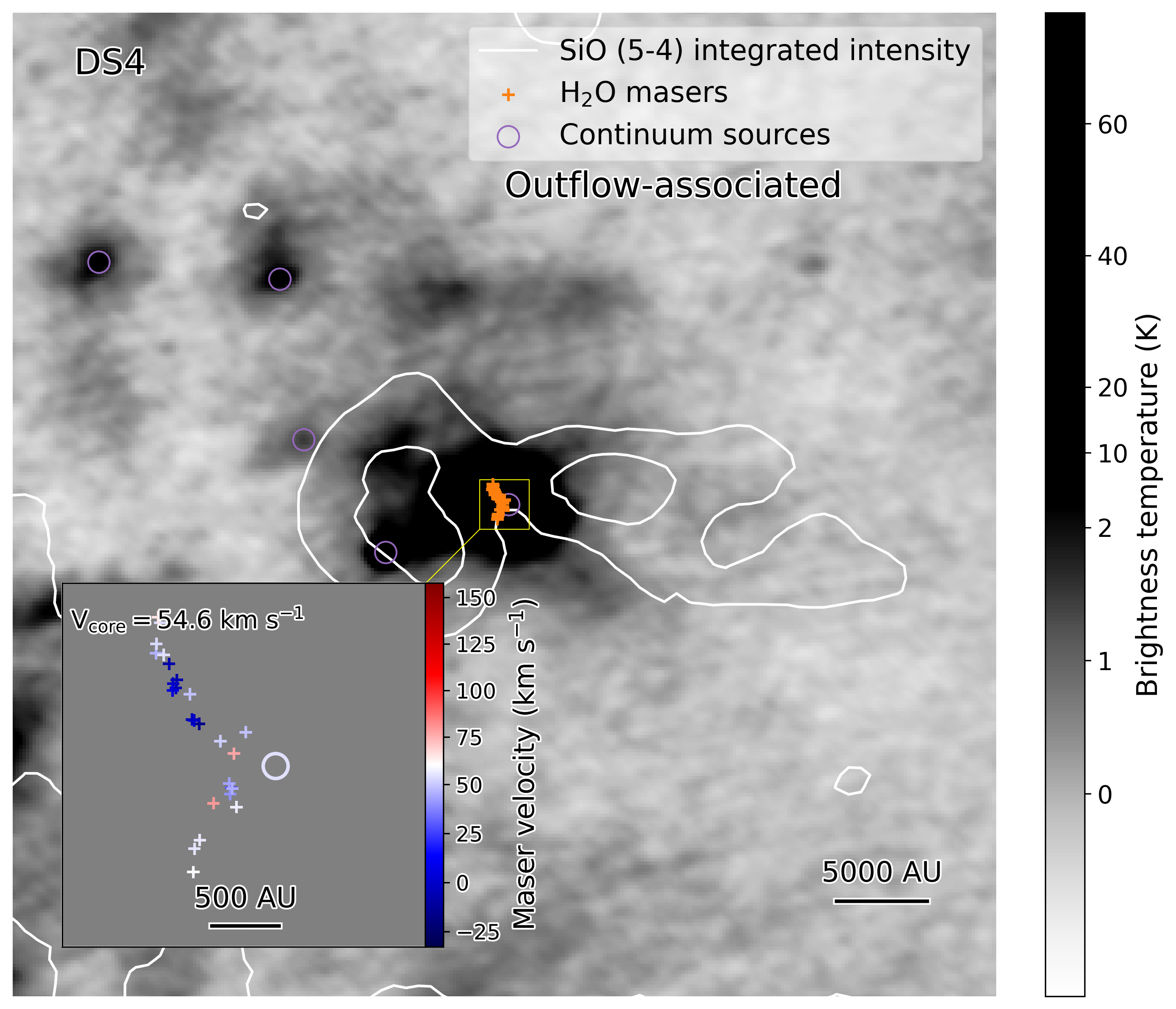}
    \end{minipage}%
    \hfill
    \begin{minipage}{0.45\textwidth} 
        \centering
        \includegraphics[width=\linewidth]{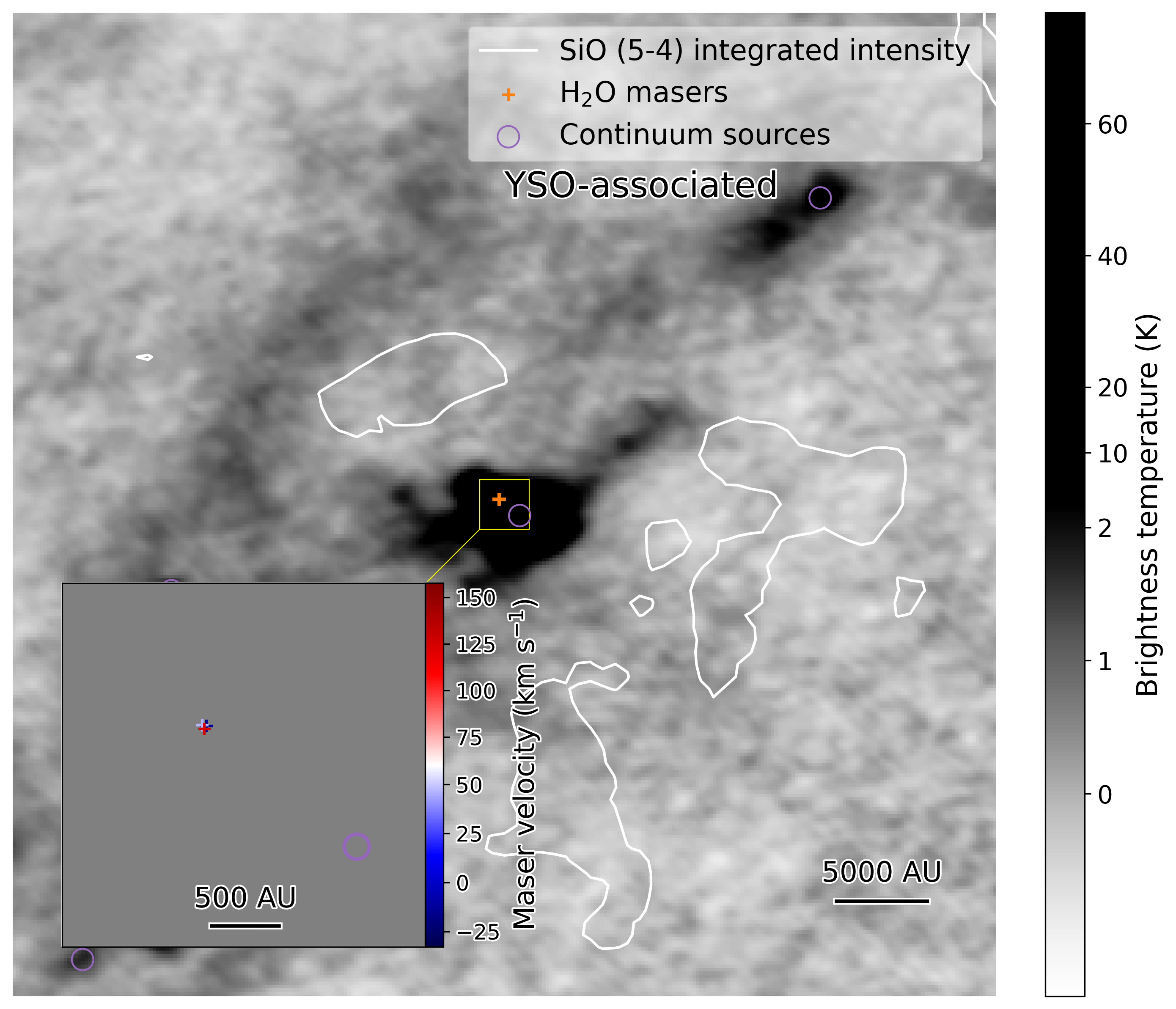}
        
    \end{minipage}
    \caption{\water masers shown on 1 mm ALMA continuum with the SiO (5-4) integrated intensity in white contours. Left: an outflow is present in SiO emission and the masers span hundreds of au. Broadly distributed masers are classified as ``outflow-associated". The masers are $<$1000 au away from the core, compared to the outflow that extends to thousands of au. The line of sight velocity of the core was reported by \cite{Jeff2024}. Right: a continuum source with no SiO emission. The tightly-packed masers with a maximum separation $<$100 au are classified as ``YSO-associated". The zoom-ins for each group of masers are available on Zenodo via  doi: \href{https://doi.org/10.5281/zenodo.15747715}{10.5281/zenodo.15747715}.
    }
    \label{fig:zoom_in_DS4}
\end{figure*}

We the use DBSCAN algorithm to cluster the \water maser features. We then classify the clusters based on their properties. If the maximum separation between members of each cluster is below 100 au, we classify all sources within this group as ``YSO-associated". Otherwise, they are classified as ``outflow-associated" (see Figure \ref{fig:zoom_in_DS4}). 
\rr{We set the minimum separation required for two groups to be considered distinct to be d$_{\mathrm{min}}$= 500 au. }Additionally, we visually inspect each cluster to identify if any group of masers that follow a larger outflow pattern were separated due to the selection parameters, but find no such cases.

We examine the spatial distribution of \water masers relative to SiO (5-4) outflows in Sgr B2 S and DS. Figure \ref{fig:zoom_in_DS4} shows one such object.
We find that the sources with structured outflows have associated masers distributed over hundreds of au, consistent with their outflow-correlated classifications.

We do not consider the location of the \water emission relative to the continuum source to perform classifications. A \water maser located close, but not on top of the flux contour, could be either from the outflow of a given continuum source or from the fainter, not-detected YSO.

Similarly, we do not consider the velocity relative to the cloud or neighboring masers. While the outflow-associated masers can reach $\sim$ 100\kms, YSO-associated masers can reach similar speeds due to the YSO's rotation or wind speed.
Assuming that masing occurs within the rotating, edge-on disk, we can expect orbital velocities as high as 150\kms for a 10\msun star at 1 au or 50\kms at 10 au.

We note that these selection criteria do not distinguish between the populations perfectly. For instance, a perfectly face-on outflow will be categorized as ``YSO-associated". A ``chain" of YSOs in a filament, below the sensitivity limit, each having a maser will be classified as an outflow. An outflow producing a few closely-spaced masers in one location would be classified as ``YSO-associated". 
Finally, if masing occurs in both the outflow and in the YSO, we classify all detections as ``outflow-associated". However, despite these possible misclassifications, the findings should be generally representative of the two populations.
The remaining sources are marked as ``ambiguous".  

\rr{We determine the number of sites of maser emission based on source clustering and cross-matching. 
We define a ``site" as a group of masers distinctly associated with one object via DBSCAN-determined clusters for YSO-associated and outflow-associated masers and via cross-matching results for the ambiguous and \hii region-associated masers. We identify 18 sites of YSO-associated masers with 53 individual detections, 33 sites of outflow-associated masers with 214 detections, 23 sites of \hii region-associated masers with 162 detections, and 69 ``ambiguous" sites. We note that 43 of the ``ambiguous" sites have a match in protostellar core catalogs. Here, ``ambiguous" only refers to whether we are able to establish the physical processes responsible for the maser emission.
The remaining 26 sites with one detection in each are unclassified, but are likely associated with protostellar cores below the previous surveys' sensitivity limits.
Finally, one maser site is associated with an evolved star. Bringing the total number of independent sites of \water maser emission to 144.
The zoom-ins for each site of maser emission, similar to those shown in Figure \ref{fig:zoom_in_DS4}, are available on Zenodo\footnote{The files are also available on GitHub: \href{https://github.com/nbudaiev/SgrB2\_VLA\_water\_masers/tree/master/zenodo\_zoomins}{https:
//github.com/nbudaiev/SgrB2\_VLA\_water\_masers/tree/
master/zenodo\_zoomins}} via doi: \href{https://doi.org/10.5281/zenodo.15747715}{10.5281/zenodo.15747715}}.

\subsubsection{Properties of different populations of masers}
\begin{figure*}[htbp] 
    \centering
    \begin{minipage}{0.45\textwidth}
        \centering
        \includegraphics[width=\linewidth]{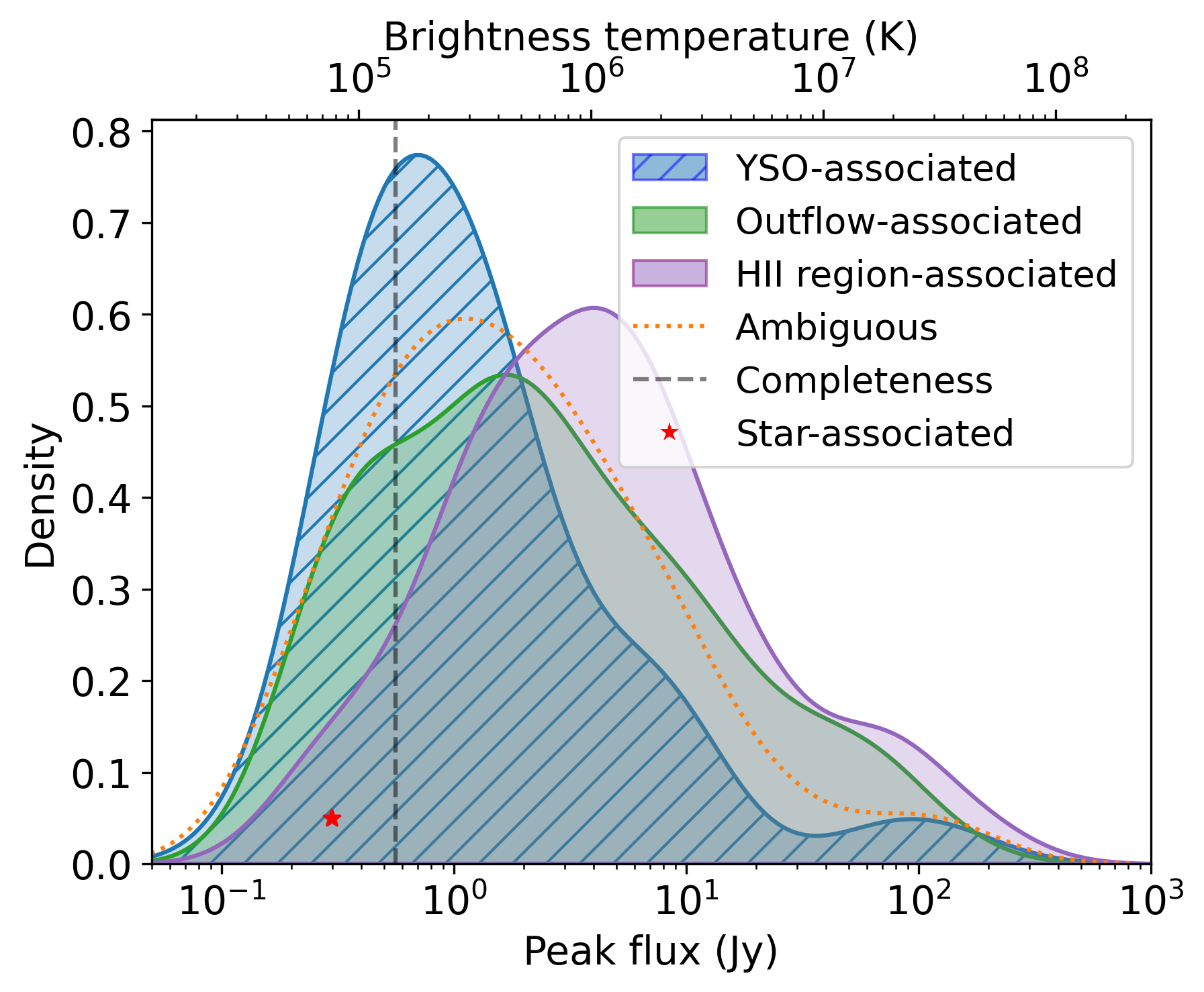}
    \end{minipage}%
    \hfill
    \begin{minipage}{0.45\textwidth} 
        \centering
        \includegraphics[width=\linewidth]{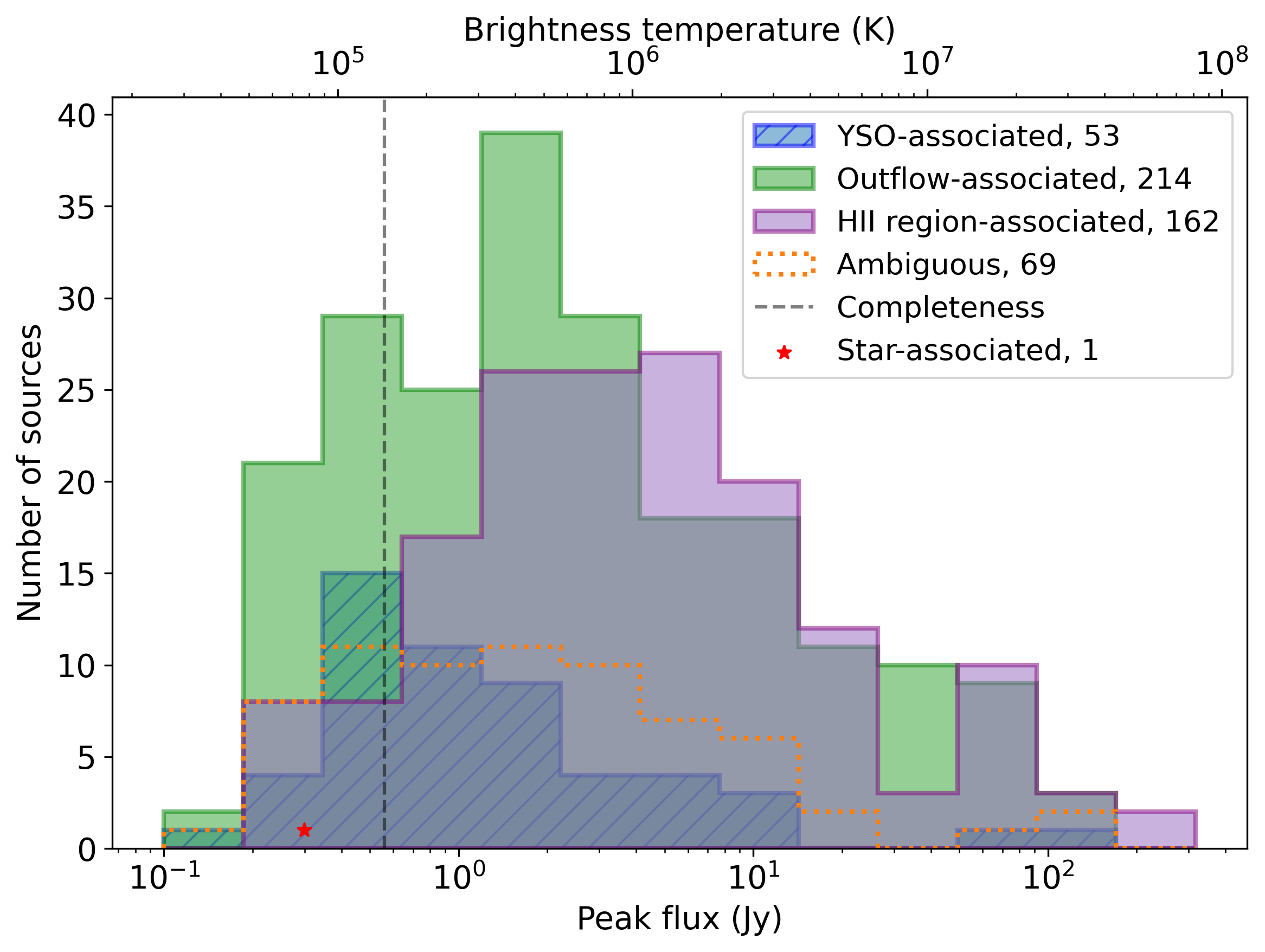}
        
    \end{minipage}
    \caption{
    Left: A kernel density estimate of \water maser brightnesses separated by their association. The \water maser associated with \hii regions are generally brighter than others. The p-value of KS test between the ``YSO-associated" and ``outflow-associated" distributions is 0.0022, indicating that the distributions are likely distinct.
    Right: Same as left, but shown as a histogram to highlight the relative abundance of each population of the masers.
    }
    \label{fig:flux_histogram}
\end{figure*}

\begin{figure}[ht]
    \centering
    \includegraphics[width=0.99\linewidth]{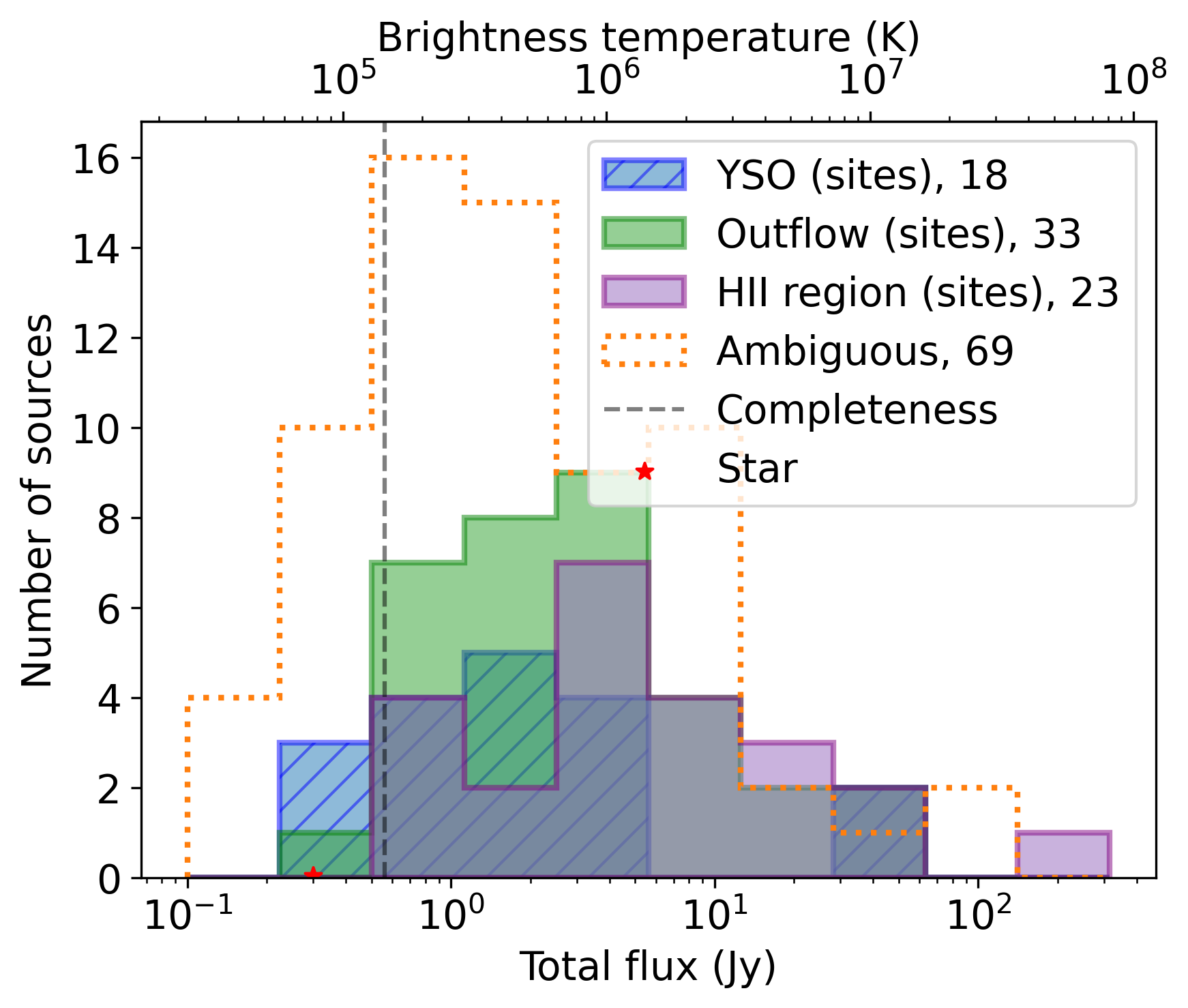}
    \caption{The histogram of the total flux in each site of maser emission. The sites are determined via source clustering for ``YSOs", ``outflows", and ``ambiguous" and by cross-matching for ``\hii regions". }
    \label{fig:flux_histogram_clusters}
\end{figure}

We show the flux distributions of the different populations of \water masers \rr{for individual detections} in Figure \ref{fig:flux_histogram} \rr{and for sites of maser emission in Figure \ref{fig:flux_histogram_clusters}.}
The flux distribution of masers associated with \hii regions peaks at $\sim$3 Jy, a factor of four above our completeness limit.
The majority of the masers associated with \hii regions are located in Sgr B2 M, with most of the remaining ones being in Sgr B2 N, as shown in Figure \ref{fig:overview_spatial}. High spatial density and the presence of extremely bright masers worsens our completeness limit within the clusters.

We perform the Kolmogorov-Smirnov (KS) test on the flux distributions of the outflow-associated and YSO-associated masers. The resulting p-value of 0.0022 indicates the  underlying distributions are a likely different between the two populations of \water masers. To our knowledge, this is the largest sample of \water masers classified by origin. 
Future, deeper observations with the new generations of radio interferometers would be needed to verify the difference in flux distributions and explore it in other environments.


We do not find any correlation between the separation between the \water maser, the associated source, and the brightness of the maser, which matches the conclusions of a more sensitive survey by \cite{Moscadelli2020}. This highlights that the absence of masers at far distances from YSOs, even with extended outflows visible in SiO, is not due to the decreasing brightness of the masers further in the outflow.



\begin{figure}
    \centering
    \includegraphics[width=0.99\linewidth]{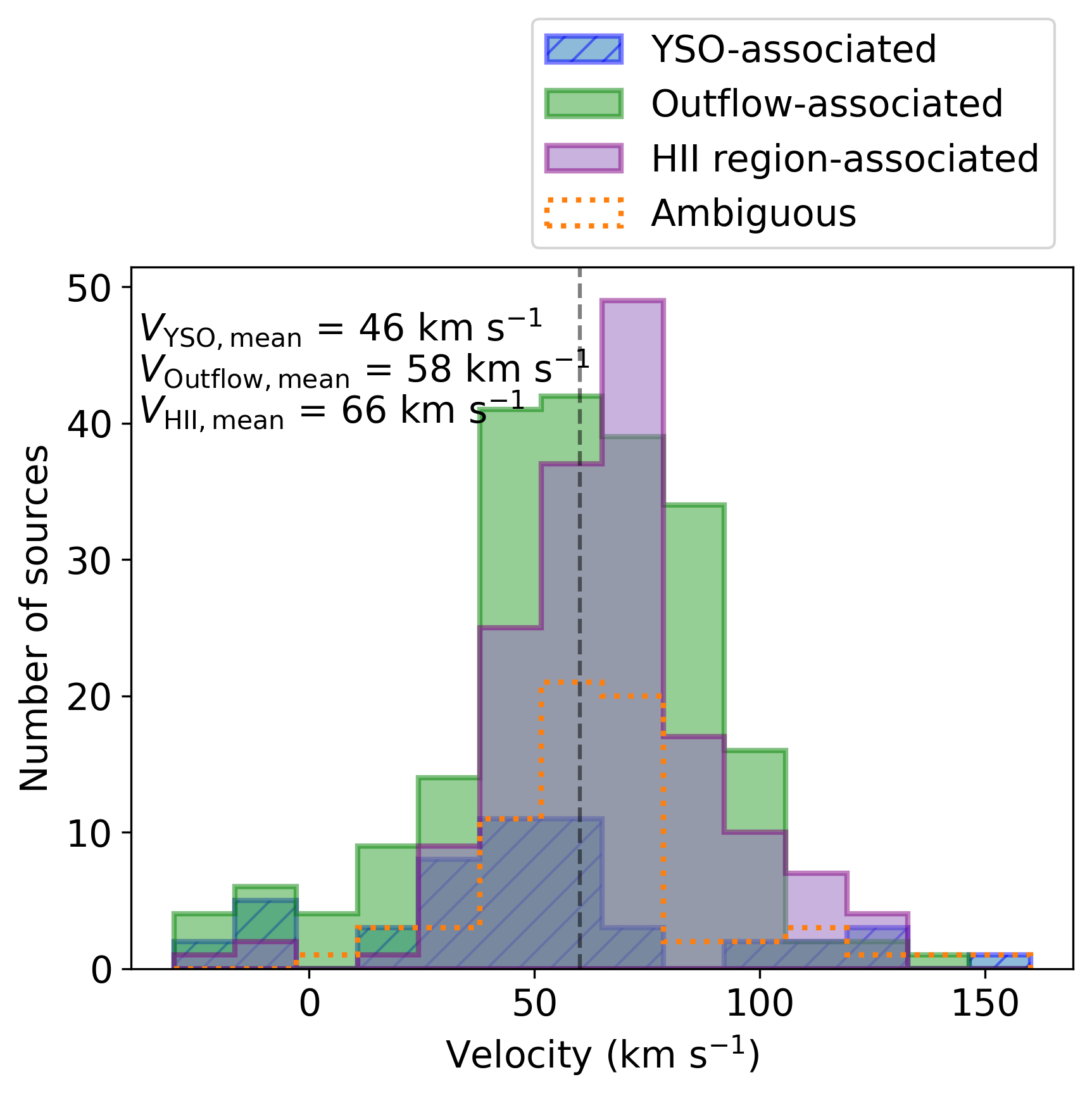}
    \caption{The line-of-sight velocity histogram of \water masers in Sgr B2 based on their origin. The velocity of the cloud is $\sim60\kms$ and is marked with a vertical dashed line. The average velocity of masers associated with \hii regions is 66\kms, which is above the velocity of the cloud. This velocity mismatch, given a large number of masers  ($>150$), indicates the unreliability of \water masers as kinematic tracers in Sgr B2.
    The YSO-associated \water masers have on-average negative velocities relative to the cloud. It is unclear what causes this trend.}
    \label{fig:velocity_histogram}
\end{figure}

\rr{We compare the observed line-of-sight velocity of the masers to their flux. We do not find any correlation across the different populations of masers. This matches previous high-resolution surveys of \water masers \citep{Moscadelli2019}.}
A histogram of \water maser \rr{velocities} broken down by population is shown in Figure \ref{fig:velocity_histogram}.
The YSO-associated masers appear to have blue-shifted velocities, whereas the outflow masers average to the mean cloud velocity.
It is unclear what is causing the velocities of the YSO-associated masers to be negative on average. Below, we discuss several possibilities.

We analyze the spatial distribution of each population of \water masers (see Figure \ref{fig:overview_spatial}).  If the majority of the YSO-associated masers are constrained to a small area of the sky, it might indicate that these sources are associated with another star-forming region along the line of sight. However, these maser detections are distributed throughout the cloud, and thus are most likely part of Sgr B2 complex.

The selection criteria was based on the spatial extent of the groups of the sources. In case our interpretation of the nature of the \water maser emission is erroneous, we still observe that tightly-clustered groups of masers have on-average negative velocities relative to the cloud. 

The distribution of the velocities of YSO-associated masers seem to lack sources at 65\kms $<$ V $<$ 90\kms. We explore whether the lack of detections are data-driven. As shown in Figure \ref{fig:noise}, the noise is the highest between $\sim50<\mathrm{V}<90\kms$ with a slight increase at around 70\kms. 
We filter the YSO-associated masers to match the average noise in the $\sim50<\mathrm{V}<90\kms$ range, $\sigma\approx0.7$\jy. The same overabundance of the slightly blue-shifted sources is present.
While it is possible that some faint sources are obscured with the higher noise, it is not enough to explain the observed trend completely.

We propose self-absorption of \water maser emission as a possible explanation for the lack of detection at $65-90\kms$. \water maser modeling by \cite{Gray2022} shows that having T$_{dust} = 100-300$ K  at gas kinetic temperatures typical for \water masing can result in absorption. If the \water maser emission comes from the central region of the YSO and is surrounded by infalling hot dust, the red-shifted emission will end up self-absorbed. Assuming a simple free-fall model for a $M = 20 \msun$ star at 100 au, a representative source from \cite{Budaiev2024}, the expected velocity is $\sim$20\kms, consistent with the gap in the data. The dust temperatures observed around hot cores in the extended Sgr B2 cloud can be as high as 500 K \citep{Bonfand2019, Jeff2024, Moller2025}. The absorption attenuates the source just enough to make it undetectable in our observations. The HII region-associated and outflow-associated \water masers will not be affected by self-absorption due to the lack of foreground high-temperature dust.

The observational parameters of this study are not ideal to determine whether the lack of $65-90\kms$ masers is due to self-absorption of \water emission. 
Measuring the separations between the YSOs and \water masers on au scales with VLBI and identifying their locations relative to the infalling gas can test the proposed self-absorption mechanism.

\subsection{Properties of the associated continuum sources}
\begin{figure}
    \centering
    \includegraphics[width=0.99\linewidth]{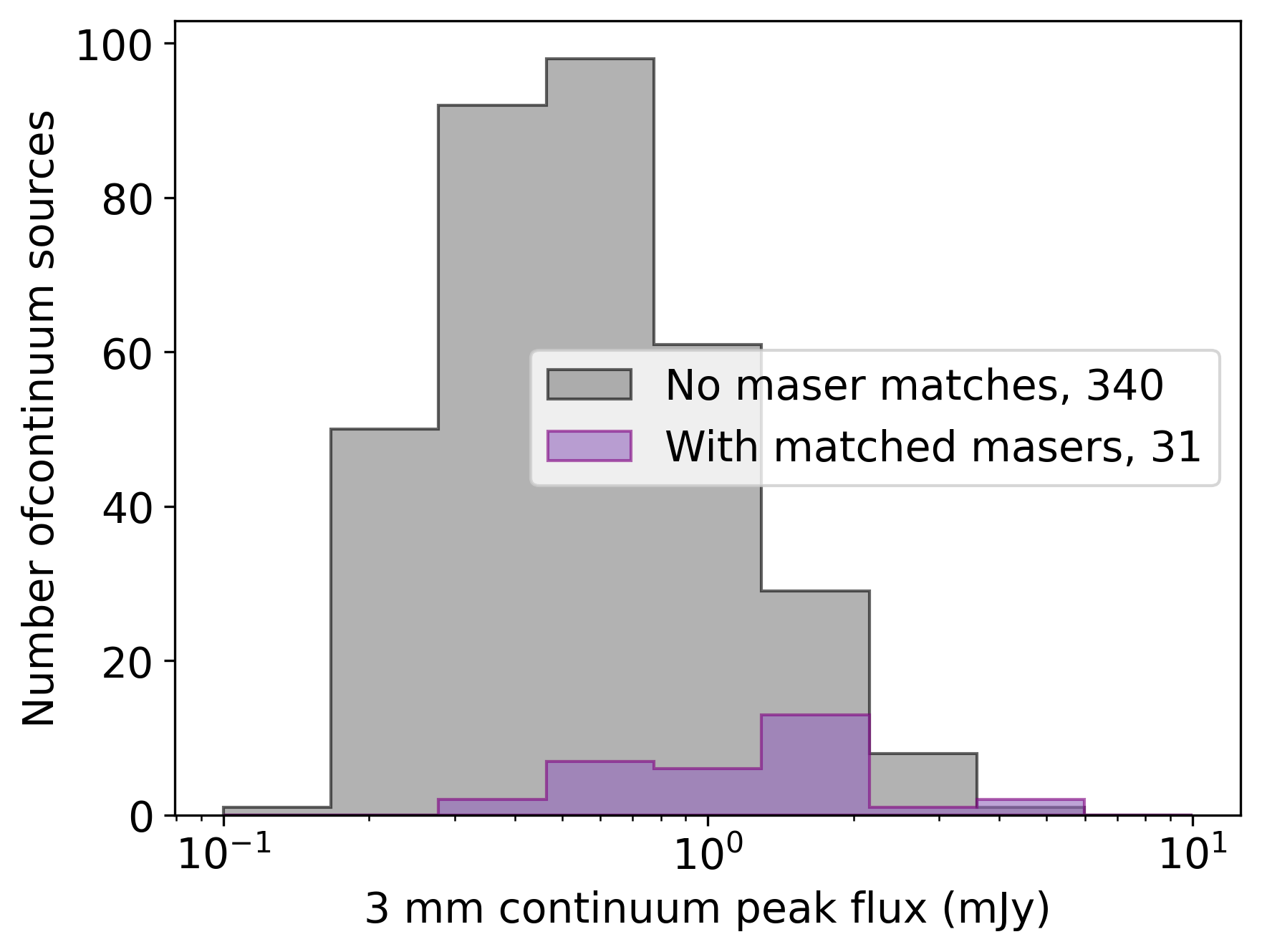}
    \caption{The flux distributions of 3 mm continuum sources with and without associated \water masers. The continuum sources with maser association are brighter on average than their counterparts. 
    }
    \label{fig:cont_hist}
\end{figure}

Using the \water maser associations from Section \ref{sec:associations}, we compare the flux distributions of 3 mm continuum cores with and without matched masers. Figure \ref{fig:cont_hist} shows that the continuum sources are brighter if they have associated \water maser emission. Of the 371 protostellar cores in Sgr B2 N and M, 31 have a corresponding \water maser, likely limited by our sensitivity. Following the conclusions of Section 3.2 from \cite{Moscadelli2020}, there exist at least twice as many masers in Sgr B2 below our sensitivity limit. 

\begin{figure}
    \centering
    \includegraphics[width=0.99\linewidth]{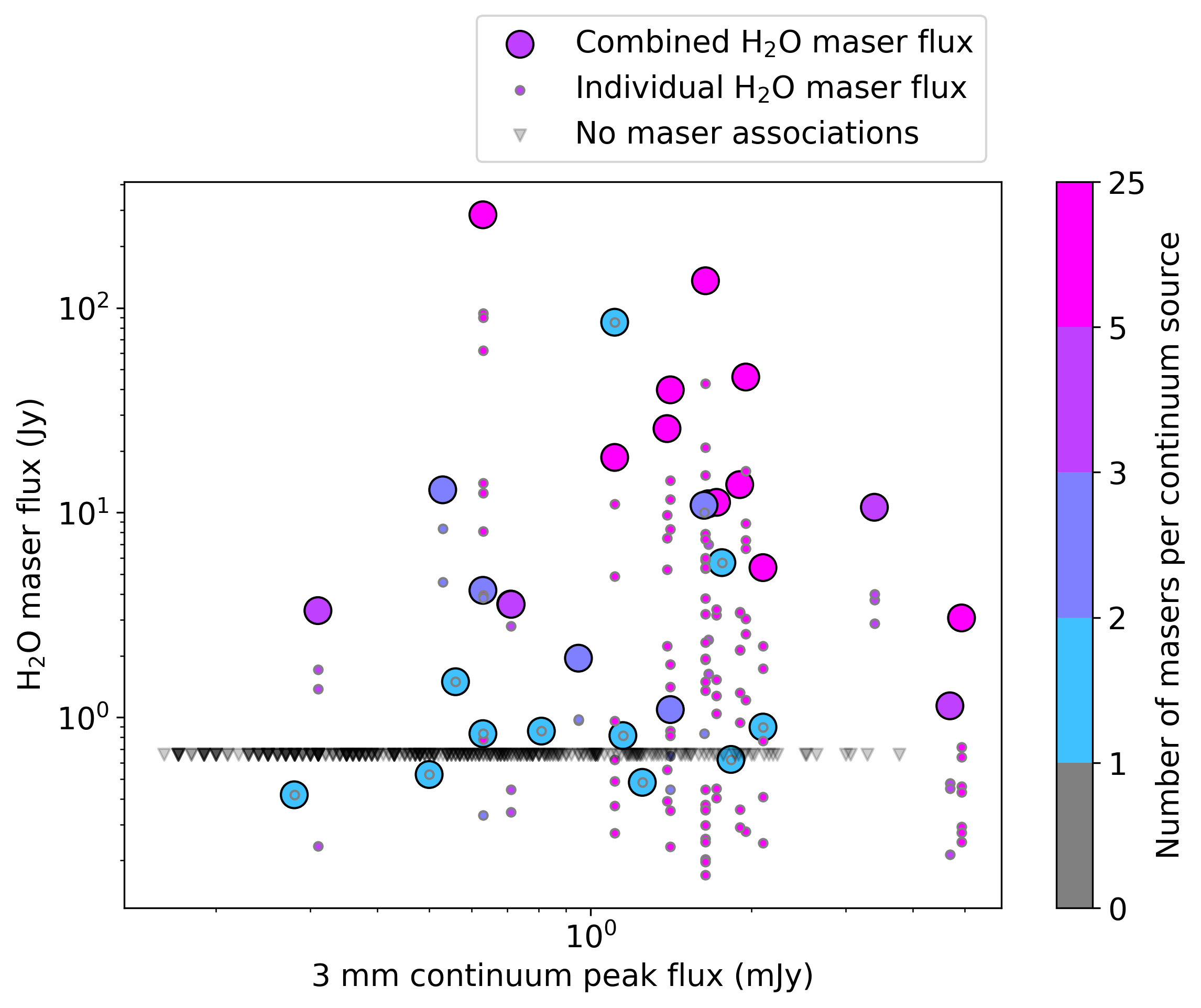}
    \caption{The 3 mm continuum flux of YSOs plotted against the flux of the associated \water masers. The color represents the number of maser detections associated with each continuum core. The small circles show the brightness of each of the individual \water masers. The cyan circles on the bottom show the brightnesses of continuum cores that do not have associated \water maser emission. 
    There is a wide range of the fractional brightnesses of individual masers from a single mm object: some are dominated by a single bright maser, some have similar brightnesses across all masers.}
    \label{fig:cont_vs_masers}
\end{figure}

We do not find a strong correlation between the 3 mm continuum flux and the \water maser luminosity as shown in Figure \ref{fig:cont_vs_masers}. 
It has been shown that lower-mass YSOs have a correlation between their bolometric luminosity and total \water maser luminosity \citep[e.g.][]{Bae2011, Urquhart2011}, and we can expect the bolometric luminosity to be proportional to the mm continuum. We attribute the lack of correlation in our observations to the spread in the ages of the star-forming clusters -- from Sgr B2 M being the oldest with the most \hii regions, and Sgr B2 DS being the youngest.

\subsubsection{Comparison to other \water maser observations}
We summarize the recent studies of \water masers in Table \ref{tab:maser_studies} to provide the context to this work. \rr{Due to the high star formation rate and relative proximity, Sgr B2 is uniquely positioned to investigate maser populations. The 144 sites of \water maser emission is close to an order of magnitude more than in any other cloud. The observed properties of the different populations of masers, can be used for further interpretation of \water maser surveys in the CMZ (e.g. D. Ward in prep.)}

\rr{
The \water maser emission in Sgr B2 would not be detected at the distance one of the most nearby star-forming galaxies: NGC 253. \cite{Gorski2019} report 13 sites of \water maser emission in NGC 253. Excluding the nuclear kilomaser W1, they observe a range of peak fluxes from 3 \mjy to 14.4 \mjy after smoothing their spectral resolution to 0.5\kms. After spatially blending all \water maser emission in Sgr B2, matching the spectral resolution, and placing at a distance of 3.5 Mpc, the observed peak flux would be $\sim$2.1 \mjy, which is just below the sensitivity of the currently available observations of NGC 253. 
}

\begin{table*}
    \centering
    \begin{tabularx}{\textwidth}{C C B B E B A}
        \hline
        \hline
        Name & Type of observations & Telescope & Resolution & Sensitivity  & $N$(maser) & $N$(site) \\
        \hline
        Sgr B2 (This work) & Star-forming cloud & VLA & 0.1\arcsec; 0.07\kms & 0.06\jy (4.1\jy kpc$^{2}$) & 499 & 144 \\
        W49 $^{(a)}$ & Star-forming cloud & KaVA (KVN + VERA) & 0.0012\arcsec; 0.42~\kms & $\sim$15\jy (1800\jy kpc$^{2}$) & $\sim$300 & 2-3$^\dag$ \\
        W51A $^{(b)}$& Star-forming cloud & VLA & 1.5\arcsec; 0.11\kms & 0.01\jy (0.3\jy kpc$^{2}$) & 999 & $>$20$^\dag$ \\
         Sgr B2 $^{(c)}$ & Star-forming cloud & VERA & 0.0015\arcsec; 0.21-~0.42\kms & $\sim$1\jy (68\jy kpc$^{2}$) &  128 & $>$20$^\dag$ \\
         \hii regions $^{(d)}$ & Survey of masers in \hii regions & KVN & 130\arcsec; 0.21\kms & 1.5\jy & 74  & 74 \\
         POETS survey $^{(e)}$ & Survey of masers in \hii regions$^*$ & VLBA & $\sim$0.001\arcsec; 0.42~\kms & $\sim$0.1\jy & 994 & 36 \\
         HOPS survey $^{(f)}$ & Survey of maser sites in the Galactic plane & ATCA & 0.4--12\arcsec; 0.42~\kms & $\sim$0.05--0.\jy & 2790 & 631 \\
         SWAG survey $^{(g)}$ & Survey of masers in the CMZ & ATCA & 30\arcsec; 0.4\kms & $\sim$0.02\jy (1.4\jy kpc$^{2}$)  & 1227 & 593 \\
        CMZ clouds $^{(h)}$ & Survey of masers in the CMZ clouds & VLA & 2.5\arcsec; 0.2\kms & 0.15\jy (10.3\jy kpc$^{2}$)  & 56 & $\sim$30 \\
         \hline
    \end{tabularx}
    \caption{Summary of recent studies of \water masers.
    $^\dag$The number of individual objects was not investigated/reported.
    $^*$\cite{Moscadelli2020} refers to the central sources as YSOs which were determined using 6 and 22\ghz continuum. In the context of this work, we refer to such objects as \hii regions.
    $^{(a)}$: \cite{Asanok2023}.
    $^{(b)}$: \cite{Zhang2024}.
    $^{(c)}$: \cite{Sakai2023}.
    $^{(d)}$: \cite{Kim2019}.
    $^{(e)}$: \cite{Moscadelli2019, Moscadelli2020}.
    $^{(f)}$: \cite{walsh2014}.
    $^{(g)}$: D. Ward (in prep.)
    $^{(h)}$: \cite{lu2019}.
    }
    \label{tab:maser_studies}
\end{table*}

\subsection{Large-scale SiO outflow from Sgr B2 N}
Due to their high masses \citep[$>10^5\msun$][]{Sanchez-Monge2017, Schworer2019}, Sgr B2 N and M are likely to evolve young massive clusters, similar to the Arches and Quintuplet. The formation and evolution of these massive clusters is still not well understood. The absence of the distributed \hii region population in Sgr B2 N provides a more transparent view of the mass flows in the cluster.

\cite{Schworer2019} reported high accretion rate (0.08–0.16 \msun \yr$^{-1}$) via eight filaments in Sgr B2 N. At the same time, \cite{Higuchi2015} see a large-scale outflow in this source. Based on SiO and SO$_2$ emission, they report an outflow rate of $\sim0.4 \msun \yr^{-1}$ on $\sim$0.1 pc scales. 

\begin{figure}[h]
    \centering
        \includegraphics[width=1\linewidth]{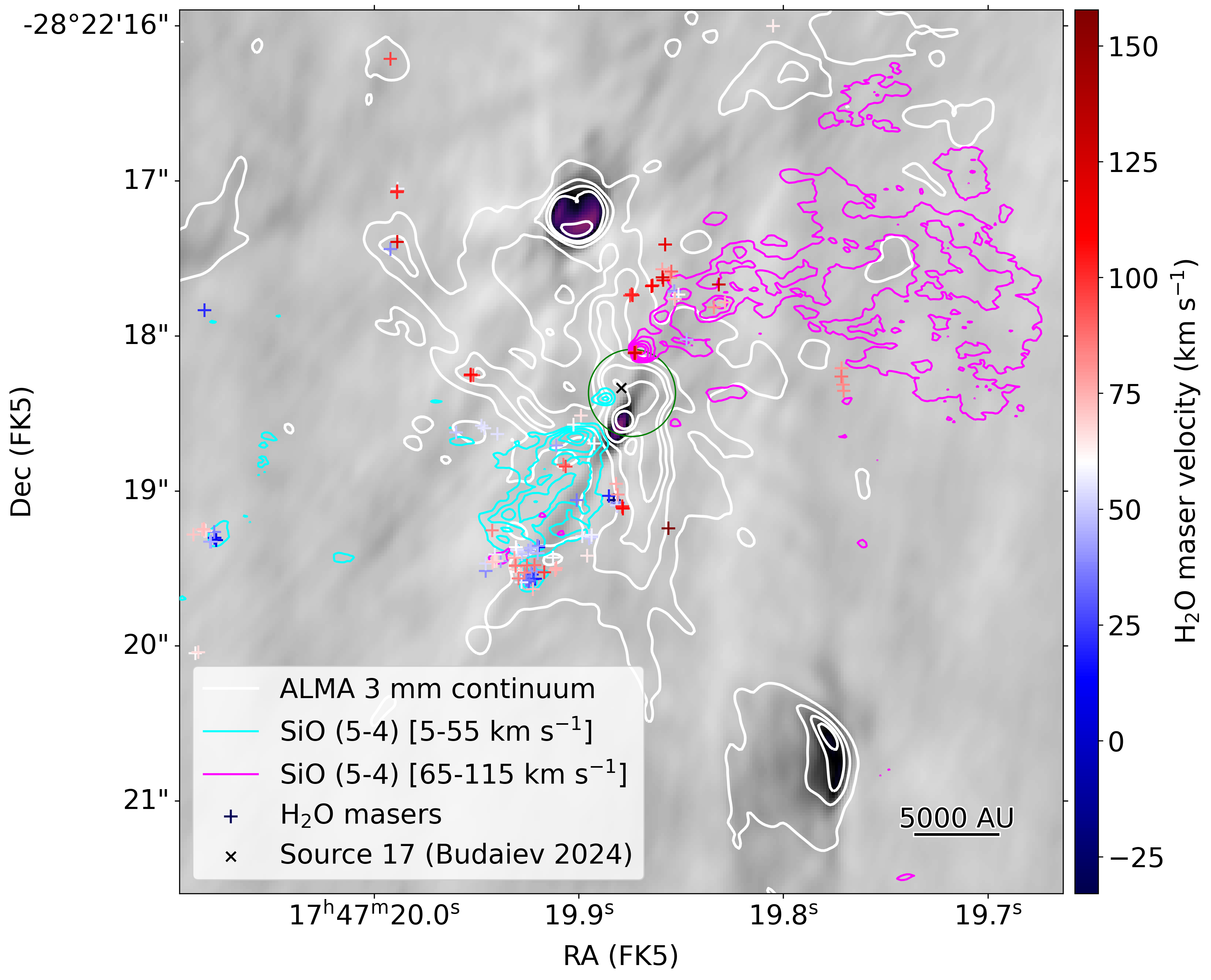}
    \caption{
    A map of the \water masers tracing the edges of the large-scale SiO outflow in Sgr B2 N.
    Colored contours show the integrated intensity (moment 0) map of SiO emission between 5 and 55 \kms (cyan) and 65 and 115\kms (magenta). These contours are plotted at 20\%, 40\%, 60\%, 80\%, and 100\% of the peak SiO (5–4) intensity. 
    The K2 region reported in previous, lower-resolution catalogs of \hii regions is shown as a green circle. The outflow appears to be aligned with source 17 from the \cite{Budaiev2024} 3 mm catalog, shown as a black x.
    ALMA 3 mm data are shown in white contours at 7,  50, 100, 150, 300, 450 K.
    The background is the 22\ghz continuum. 
    }
    \label{fig:n_outflow}
\end{figure}

We show the structure of the outflow in Sgr B2 N on the scale of thousands of aus in Figure \ref{fig:n_outflow}, as seen in SiO (ALMA PID: 2016.1.00550.S). 
\rr{The outflow boundaries are neighboring and likely interacting with the filaments that are feeding the proto-cluster. The blue lobe is located between filaments F01-F02 and F03-F04 and the red lobe is located between filaments F07 and F08 defined in \cite{Schworer2019}. This orientation aligns with the velocities of the filaments. Such an inflow of material could be responsible for the collimation of the outflow. 
The \water\ masers, which trace shocked gas, are aligned with the edges of the outflow as seen in SiO emission. This alignment suggests that we are observing the actual physical boundaries of the outflow, unaffected by limitations such as resolution, sensitivity, projection, or optical depth effects.
A similar configuration is observed in the massive protostellar outflow W51-North, where \water\ masers are excited at the leading edges of compact SiO outflow lobes \citep[][see their Figure 7]{Goddi2020}, supporting the interpretation that masers trace shock fronts at the outflow interface.
The ``outflow" could therefore be a combination of ionization, radiation, and smaller-scale, isotropic outflows within the cluster that are then shaped to appear as a bipolar outflow \citep{Peters2012}.}

\rr{However, the high-resolution data suggest an alternative possibility. The outflow is typically attributed to Sgr B2 N K2 source (e.g. \citealt{Higuchi2015}, \citealt{Schmiedeke2016}). 
The $\sim$500 au resolution mm continuum catalog shows that Sgr B2 N K2 source splits into three \hii regions \citep{Budaiev2024}. Visually, there is only a single source that lies along the outflow axis and thus could be responsible for the outflow: source 17 (see Figure \ref{fig:n_outflow}). 
While identifying the exact nature of this outflow is beyond the scope of this work,
it is possible that this massive outflow is produced by one, or at most two, accreting objects.} The previously reported properties \rr{of the outflow} are consistent with those at the early stages of massive star formation \citep{Zapata2010}.

\subsection{Maser-based proper motions measurements}\label{sec:PM_Reid}
Currently, the most cited measurements of the proper motion of Sgr B2 come from \cite{Reid2009}. The authors used VLBI observations of two \water masers, both in the hearts of Sgr B2 N and M, and reported an average motion in Galactic longitude of -4.05 mas yr$^{-1}$ ($\sim$160\kms). They acknowledged that the measurement does not account for the internal motion of the masers and thus assign a probable uncertainty of 1 mas y$^{-1}$ ($\sim$40\kms). However, as shown in this work, this uncertainty is significantly underestimated (see Figure \ref{fig:overview_new_detections}). As discussed in Section \ref{sec:nature}, depending on the pumping mechanism the masers can have relative velocities of over 100\kms. We detect masers with relative velocities of up to 100\kms and are possibly observationally limited. As seen in Figure \ref{fig:velocity_vs_dec}, there are detected sources up to the edges of the observed frequencies.
Furthermore, sources in the vicinity of a massive cluster are subject to high gravitational potentials and strong feedback, which can further impact the proper motion measurements. 
For example, when measuring the velocity dispersion in Sgr B2 via radio recombination lines, \cite{Ginsburg2018CFE} reports a group of \hii regions with $v_{rel} = \sim$28\kms.

\cite{Sakai2023} measured proper motions of 55 \water masers in Sgr B2 M. They report internal proper motions up to 3.5 mas \yr$^{-1}$ ($\sim$140 \kms), further highlighting the high uncertainty of using a small sample of \water masers to measure proper motions. 

In conclusion, one must take great care when estimating the proper motion of a massive molecular cloud like Sgr B2 using masers with multiple pumping mechanisms. As has been mentioned in the sections above, even with a large sample, it is possible to introduce a large bias in the measurement. For example, if there exists a cloud with a density gradient, which results in \hii regions facing less pressure on one side, such \hii regions will have more \water masers in the denser side of the expanding shell. If all of the \hii regions are still expanding, measuring their proper motions using \water masers will introduce a large velocity bias in the direction of increasing density. 

Thus, if \water masers are used to measure the proper motions of the cloud, an analysis of the nature of the masers, and thus the possible biases, should be undertaken. 
\rr{
Follow up VLA observations are likely to face issues similar to those described in this work: complicated phase calibration, confusion due to source density and maser variability, and non-standard imaging procedures. Thus, very long baseline observation are more suitable for the studies of maser proper motions. Describing the proper motions within the cloud is one of the key elements to understanding the formation history of the cloud and can provide insights into the 3D kinematics of the CMZ.
}
Alternatively, the dusty cores themselves can be used to measure the proper motions in the CMZ. High-resolution ALMA observations in the Galactic Center began almost ten years ago, allowing for tens of \kms velocity resolution.

\section{Conclusions}\label{sec:conclusions}
We presented a catalog of 499 \water masers in Sgr B2, the most star-forming cloud in the CMZ. We classified the nature of the masers by cross-matching with catalogs of \hii regions, protostellar cores, and massive stars. 
At this time, Sgr B2 is the only location where we can study population-level properties of \water masers produced by different star-forming objects within the same environment. We identified 144 unique sites of \water maser emission, almost an order of magnitude more than in any other cloud.
Of those, 23 are associated with \hii regions, 33 with protostellar outflows, and 18 with YSOs. 
\begin{itemize}
    \item The \water maser produced by YSO outflows extend only up to thousand au scales, compared to the SiO emission which extends up to tens of thousands au. 
    \item 3 mm continuum emission from YSOs is generally brighter if \water maser emission is present. However, there is no strong correlation between the brightness of masers and the associated mm continuum.
    \item The YSO-associated and outflow-associated \water masers seem to have different observable properties: the YSO-associated \water masers are slightly dimmer; in addition, there is an unexplained tendency for YSO-associated masers to have negative relative velocities on average.
    \item \water masers trace the colliding material flows in the center of Sgr B2 N. Combined with high-resolution SiO observations, the large-scale outflow appears to originate from a single star-forming object.
    \item If \water masers are used to measure the proper motion of a massive cloud, the nature of the water masers should be established to reduce the uncertainty. 
\end{itemize}

\begin{acknowledgements}
\rr{We thank Sheila Sagear and Cara Battersby for helpful discussions about this work.}
The National Radio Astronomy Observatory is a facility of the National Science Foundation operated under cooperative agreement by Associated Universities, Inc.
This paper makes use of the following ALMA data: ADS/JAO.ALMA\#2016.1.00550.S, ADS/JAO.ALMA\#2017.1.00114. ALMA is a partnership of ESO (representing its member states), NSF (USA) and NINS (Japan), together with NRC (Canada), MOST and ASIAA (Taiwan), and KASI (Republic of Korea), in cooperation with the Republic of Chile. The Joint ALMA Observatory is operated by ESO, AUI/NRAO and NAOJ. 
A.G. and N.B. acknowledge support from the NSF under AST 2206511.
A.G. acknowledges support from NSF under CAREER 2142300 and AST 2206511.
C.G. acknowledges financial support by 
the European Union NextGenerationEU RRF M4C2 1.1 project n. 2022YAPMJH and 
FAPESP (Funda\c{c}\~ao de Amparo \'a Pesquisa do Estado de S\~ao Paulo) under grant 2021/01183-8. 
The authors acknowledge University of Florida Research Computing for providing computational resources and support that have contributed to the research results reported in this publication. URL: http://www.rc.ufl.edu.
\end{acknowledgements}

\begin{contribution}

N.B. led the analysis, writing, and interpretation. A.G. acquired the observations, oversaw the project progress, and contributed to the interpretation. C.G., A.S-M., and A.S. were instrumental in acquiring the observations and contributed to the interpretation. D.J. contributed to the interpretation. P.S. and C.D.P. were instrumental in acquiring the observations.

\end{contribution}

\facilities{VLA, ALMA}

\software{
\texttt{astropy} \citep{astropy:2013, astropy:2018},  \texttt{pyspeckit} \citep{pyspeckit}, \texttt{astrodendro} \citep{Rosolowsky2008}, CASA version version 6.6.0.2 \citep{McMullin2007}, \texttt{scikit-image} \citep{scikit-image}.
  }

\vspace{5mm}

\appendix

\section{Catalog of \water masers in Sgr B2}\label{app:catalog}

We present an excerpt from the catalog of \water masers in Sgr B2 in Table \ref{tab:catalog}. The full catalog is available in machine-readable format.

\begin{deluxetable}{cccccccccccccc}
\tablecaption{First 35 entries of the \water maser catalog in Sgr B2. \label{tab:catalog}}
\tablehead{
ID & RA & Dec & RA$_e$ & Dec$_e$ & Flux & Velocity & Sigma & Fit success & Matched catalog & Match ID & Matched source & Site type & Site ID \\
 & \degree & \degree & $\mathrm{{}^{\prime\prime}}$ & $\mathrm{{}^{\prime\prime}}$ & \jy & \kms & \kms &  &  &  &  &  &  \\}

\startdata
0 & 266.832 & -28.384 & 0.0012 & 0.0025 & 0.55676 & -29.35 & 0.076 & TRUE & B24 & 178 & core & YSO & 1 \\
1 & 266.832 & -28.384 & 0.0003 & 0.0007 & 5.2794 & -25.77 & 0.046 & TRUE & B24 & 178 & core & YSO & 1 \\
2 & 266.833 & -28.372 & 0.0004 & 0.0008 & 1.85667 & -23.19 & 0.025 & TRUE & No match & No match & No match & Outflow & 19 \\
3 & 266.834 & -28.384 & 0.0023 & 0.005 & 0.35604 & -17.31 & 0.036 & TRUE & B24 & 19 & HII & HII & 56 \\
4 & 266.834 & -28.384 & 0.0014 & 0.0029 & 0.44133 & -14.71 & 0.033 & TRUE & B24 & 19 & HII & HII & 56 \\
5 & 266.834 & -28.384 & 0.0015 & 0.0031 & 0.51742 & -13.32 & 0.041 & TRUE & B24 & 19 & HII & HII & 56 \\
6 & 266.832 & -28.384 & 0.0004 & 0.0008 & 2.23226 & -10.52 & 0.051 & TRUE & B24 & 178 & core & YSO & 1 \\
7 & 266.832 & -28.384 & 0.0003 & 0.0006 & 9.76375 & -6.4 & 0.042 & TRUE & B24 & 178 & core & YSO & 1 \\
8 & 266.832 & -28.386 & 0.0033 & 0.0063 & 0.29287 & -5.37 & 0.028 & TRUE & B24 & 168 & core & Outflow & 20 \\
9 & 266.83 & -28.385 & 0.0021 & 0.0044 & 0.62262 & -4.62 & 0.023 & TRUE & B24 & 215 & core & YSO & 2 \\
10 & 266.832 & -28.384 & 0.0002 & 0.0005 & 7.5093 & -4.65 & 0.036 & TRUE & B24 & 178 & core & YSO & 1 \\
11 & 266.834 & -28.372 & 0.0005 & 0.001 & 1.71108 & 7.0 & 0.021 & TRUE & No match & No match & No match & Outflow & 21 \\
12 & 266.844 & -28.375 & 0.0022 & 0.0043 & 0.29912 & 9.61 & 0.033 & TRUE & No match & No match & star & Star & 75 \\
13 & 266.834 & -28.372 & 0.0012 & 0.0024 & 0.4292 & 10.86 & 0.036 & TRUE & No match & No match & No match & Outflow & 21 \\
14 & 266.829 & -28.385 & 0.0018 & 0.004 & 0.23549 & 13.16 & 0.035 & TRUE & B24 & 227 & core & Outflow & 22 \\
15 & 266.833 & -28.372 & 0.0004 & 0.0009 & 8.60395 & 15.64 & 0.015 & TRUE & B24 & 4 & core & Outflow & 23 \\
16 & 266.833 & -28.372 & 0.0012 & 0.0028 & 1.34287 & 16.37 & 0.024 & TRUE & B24 & 4 & core & Outflow & 23 \\
17 & 266.828 & -28.384 & 0.0016 & 0.0035 & 0.4187 & 16.92 & 0.032 & TRUE & B24 & 229 & core & Ambiguous & 76 \\
18 & 266.833 & -28.372 & 0.002 & 0.0037 & 0.45586 & 18.17 & 0.034 & TRUE & B24 & 4 & core & Outflow & 23 \\
19 & 266.832 & -28.386 & 0.0004 & 0.0009 & 2.60437 & 19.01 & 0.029 & TRUE & B24 & 167 & core & YSO & 3 \\
20 & 266.833 & -28.372 & 0.0015 & 0.0036 & 0.82865 & 19.56 & 0.022 & TRUE & B24 & 4 & core & Outflow & 23 \\
21 & 266.833 & -28.372 & 0.0012 & 0.0025 & 0.41599 & 20.78 & 0.023 & TRUE & No match & No match & No match & Outflow & 19 \\
22 & 266.833 & -28.372 & 0.0017 & 0.0035 & 1.38237 & 21.93 & 0.053 & TRUE & B24 & 4 & core & Outflow & 23 \\
23 & 266.834 & -28.372 & 0.0005 & 0.001 & 3.08462 & 22.04 & 0.015 & TRUE & No match & No match & No match & Ambiguous & 77 \\
24 & 266.834 & -28.384 & 0.0029 & 0.0051 & 0.59177 & 22.15 & 0.061 & TRUE & B24 & 21 & HII & HII & 57 \\
25 & 266.832 & -28.386 & 0.0002 & 0.0005 & 8.82692 & 22.34 & 0.037 & TRUE & B24 & 167 & core & YSO & 3 \\
26 & 266.832 & -28.386 & 0.0062 & 0.0138 & 0.46005 & 23.04 & 0.028 & TRUE & B24 & 168 & core & Ambiguous & 78 \\
27 & 266.829 & -28.384 & 0.0019 & 0.004 & 0.46551 & 24.27 & 0.025 & TRUE & No match & No match & No match & YSO & 4 \\
28 & 266.833 & -28.372 & 0.0004 & 0.0009 & 1.75948 & 24.38 & 0.022 & TRUE & B24 & 4 & core & Outflow & 23 \\
29 & 266.832 & -28.386 & 0.0017 & 0.0037 & 0.64469 & 25.3 & 0.024 & TRUE & B24 & 167 & core & YSO & 3 \\
30 & 266.833 & -28.371 & 0.0016 & 0.0033 & 0.77047 & 25.53 & 0.028 & TRUE & B24 & 159 & core & YSO & 5 \\
31 & 266.829 & -28.384 & 0.0008 & 0.0016 & 1.67536 & 29.01 & 0.02 & TRUE & No match & No match & No match & YSO & 4 \\
32 & 266.833 & -28.372 & 0.0003 & 0.0006 & 20.25269 & 28.94 & 0.047 & TRUE & B24 & 4 & core & Outflow & 23 \\
33 & 266.834 & -28.385 & 0.0038 & 0.0075 & 0.78611 & 29.23 & 0.024 & TRUE & B24 & 16 & HII & HII & 53 \\
34 & 266.834 & -28.384 & 0.0032 & 0.007 & 0.24507 & 29.72 & 0.044 & TRUE & B24 & 27 & HII & HII & 62 \\
\enddata
\end{deluxetable}

\section{Source extraction with \textit{Duchamp}}\label{sec:duchamp}
We tested source extraction software \texttt{DUCHAMP} \citep{duchamp}. While this software is more appropriate when working with large 3D cubes, we find that it does not behave well with clustered groups of sources. We were unable to de-blend most of the detections in Sgr B2 N and M. 

\section{VLA 22\ghz continuum}\label{app:continuum_image}
\begin{figure}
    \centering
    \includegraphics[width=1\linewidth]{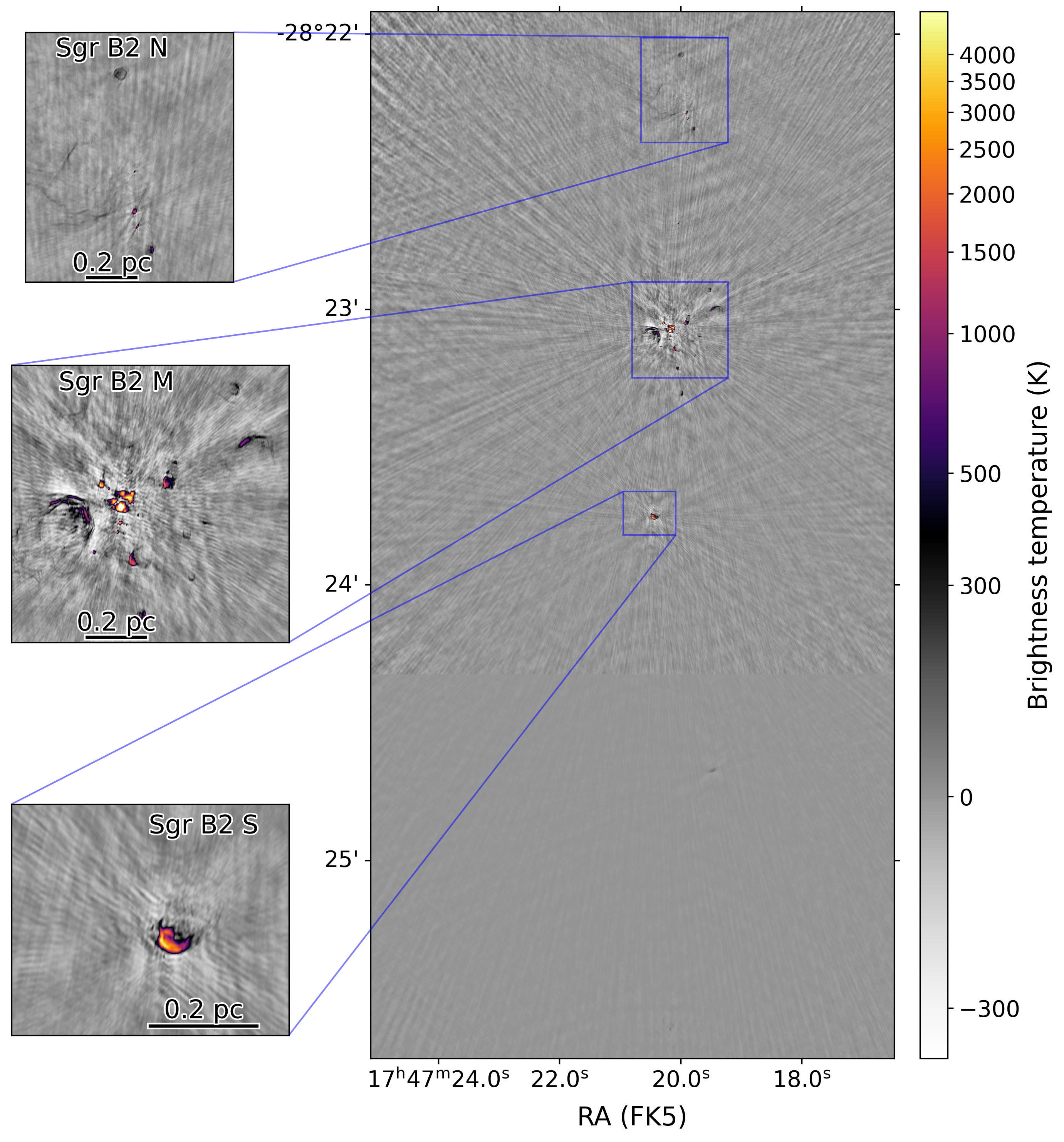}
    \caption{22\ghz VLA continuum image of Sgr B2. The fits file is available on Zenodo via doi: \href{https://doi.org/10.5281/zenodo.15747715}{10.5281/zenodo.15747715}.
    }
    \label{fig:continuum_22GHz}
\end{figure}
We include the 22\ghz continuum map used to align the VLA observations to ALMA data. The map is shown in Figure \ref{fig:continuum_22GHz}. The reduced fits files for each pointing is available on Zenodo via doi: \href{https://doi.org/10.5281/zenodo.15747715}{10.5281/zenodo.15747715}.


\bibliography{references}{}
\bibliographystyle{aasjournalv7}

\end{document}

%% file: LaTeX-macros.tex
\newcommand{\y}[1]{\ensuremath{y_{#1}}\xspace}

\newcommand{\kms}{\ensuremath{\,{\rm km\,s^{-1}}}\xspace}

\newcommand{\m}{\ensuremath{\,{\rm m}}\xspace}

\newcommand{\pc}{\ensuremath{\,{\rm pc}}\xspace}
\newcommand{\kpc}{\ensuremath{\,{\rm kpc}}\xspace}
\newcommand{\K}{\ensuremath{\,{\rm K}}\xspace}

\newcommand{\yr}{\ensuremath{\,{\rm yr}}\xspace}

\newcommand{\ghz}{\ensuremath{\,{\rm GHz}}\xspace}

\newcommand{\degree}{\ensuremath{\,^\circ}\xspace}

\newcommand{\jy}{\,Jy\xspace}
\newcommand{\jyb}{\ensuremath{{\rm \,Jy\,beam^{-1}}}\xspace}
\newcommand{\mjy}{\,mJy\xspace}
\newcommand{\mjyb}{\ensuremath{{\rm \,mJy\,beam^{-1}}}\xspace}

\newcommand{\msun}{\ensuremath{{\rm \,M_\odot}}\xspace}     
 
\newcommand{\h}[1]{\ensuremath{^{#1}{\rm H}}\xspace}

\newcommand{\hii}{{\rm H\,{\small II}}\xspace}

\newcommand{\sio}{\ensuremath{\rm SiO}\xspace}

\newcommand{\water}{\ensuremath{\rm H_2O}\xspace}